\newif\ifonecol
\onecoltrue

\ifonecol
    \documentclass[12pt,draftclsnofoot,onecolumn]{IEEEtran}
\else
    \documentclass[journal,comsoc]{IEEEtran}
\fi

\usepackage{amsmath,graphics,amssymb,epsfig,subfigure,color,amsthm,cite}
\usepackage{array}
\usepackage{multirow}
\usepackage{enumerate}
\usepackage{algorithmic}
\usepackage{algorithm}
\usepackage{bm}
\usepackage{float}

\newtheorem{thm}{Theorem}
\newtheorem{lem}{Lemma}
\newtheorem{cor}{Corollary}

\newcommand{\argmax}{\operatornamewithlimits{argmax}}
\newcommand{\argmin}{\operatornamewithlimits{argmin}}
\makeatletter
\newcommand{\vast}{\bBigg@{4.5}}
\newcommand{\Vast}{\bBigg@{7.5}}
\makeatother

\begin{document}
\ifonecol
	\title{\LARGE{Robust Data Detection for MIMO Systems with One-Bit ADCs: A Reinforcement Learning Approach}}
\else
	\title{Robust Data Detection for MIMO Systems with One-Bit ADCs: A Reinforcement Learning Approach}
\fi

\author{Yo-Seb Jeon, Namyoon Lee, and H. Vincent Poor
	\thanks{Y.-S. Jeon is with the Department of Electrical Engineering, Princeton University, Princeton, NJ 08544. He is also with the Department of Electrical Engineering, POSTECH, Pohang, Gyeongbuk 37673, South Korea (e-mail: yoseb.jeon@postech.ac.kr).}
	\thanks{N. Lee is with the Department of Electrical Engineering, POSTECH, Pohang, Gyeongbuk 37673, South Korea (e-mail: nylee@postech.ac.kr).}
	\thanks{H. V. Poor is with the Department of Electrical Engineering, Princeton University, Princeton, NJ 08544 (e-mail: poor@princeton.edu).}
}
\vspace{-2mm}	

\maketitle
\vspace{-12mm}

\begin{abstract} 
	The use of one-bit analog-to-digital converters (ADCs) at a receiver is a power-efficient solution for future wireless systems operating with a large signal bandwidth and/or a massive number of receive radio frequency chains. This solution, however, induces a high channel estimation error and therefore makes it difficult to perform the optimal data detection that requires perfect knowledge of likelihood functions at the receiver. In this paper, we propose a likelihood function learning method for multiple-input multiple-output (MIMO) systems with one-bit ADCs using a reinforcement learning approach. The key idea is to exploit input-output samples obtained from data detection, to compensate the mismatch in the likelihood function. The underlying difficulty of this idea is a label uncertainty in the samples caused by a data detection error. To resolve this problem, we define a Markov decision process (MDP) to maximize the accuracy of the likelihood function learned from the samples. We then develop a reinforcement learning algorithm that efficiently finds the optimal policy by approximating the transition function and the optimal state of the MDP. Simulation results demonstrate that the proposed method provides significant performance gains for the optimal data detection methods that suffer from the mismatch in the likelihood function.
\end{abstract}


\section{Introduction}\label{Sec:Intro}
Future wireless systems are expected to use a large signal bandwidth and/or a massive antenna array to achieve high data rates beyond hundreds of Gbits/sec \cite{Swindle:14,Sun:12,Han:15}. Unfortunately, implementing these techniques also brings a significant power consumption problem at a receiver. The main cause of this problem is a high-precision (12$\sim$16 bits) analog-to-digital converter (ADC) employed per radio frequency (RF) chain at the receiver, because its power consumption increases linearly with both the signal bandwidth (i.e., the sampling rate) and the number of receive RF chains \cite{Murmann,Walden:99,Singh:09,Orhan:09}. To resolve this problem, the use of one-bit ADCs has received a great deal of attention \cite{Mezghani:07,Madhow:09,Bjornson:15,Hyowon:18,Jeon:arXiv:18,Wang:14,Choi:16,Hong:18,Jeon:TWC:18,Hong:CL:18,He:18,Studer:16,Li:2017,Lok:98,Mezghani:10,Stockle:16,Mo:18,Wen:16,Wang:17,Sun:18,Jeon:TVT:18}. This solution provides an exponential reduction in the power consumption of the ADCs and therefore makes it possible to compensate the power increase in the future wireless systems.

In multiple-input multiple-output (MIMO) systems operating with one-bit ADCs, perfect knowledge of likelihood functions is necessary at the receiver to perform the optimal data detection \cite{Wang:14,Choi:16,Hyowon:18,Jeon:arXiv:18}. The most common approach to attain this knowledge is to compute the likelihood functions based on channel information obtained via pilot-assisted channel estimation methods (e.g., \cite{Choi:16,Studer:16,Li:2017,Lok:98,Mezghani:10,Stockle:16,Mo:18}). This channel information, however, is inaccurate when employing one-bit ADCs because coarse quantization at the ADCs fundamentally limits the available information at the receiver. Such inaccurate channel information causes a mismatch in the likelihood functions computed at the receiver and therefore results in a severe performance degradation in the optimal data detection methods. To facilitate reliable communication in the MIMO systems with one-bit ADCs, it is essential to design a likelihood function estimation method that is robust to the mismatch caused by the inaccurate channel information.

In our prior works \cite{Jeon:WCNC:18,Jeon:ICC:18}, we made the first attempt to use a reinforcement learning approach to design a likelihood function learning method for MIMO systems with one-bit ADCs. A major limitation of the methods in \cite{Jeon:WCNC:18,Jeon:ICC:18} is that they exploit only a part of input-output samples obtained from the data detection, while being applicable to specific data detection methods. In this paper, we make substantial progress toward this direction; we define a Markov decision process (MDP) that optimizes the exploitation of all the input-output samples to maximize the accuracy of the learned likelihood function, and then develop a practical algorithm to efficiently solve the MDP. We also improve the applicability of the learning method, so that it is universally applicable to any data detection method that utilizes the likelihood functions as the sufficient statistics in the MIMO systems with one-bit ADCs.

\subsection{Related Work}
Data detection methods for MIMO systems with one-bit ADCs have been intensively studied in the literature \cite{Hyowon:18,Jeon:arXiv:18,Wang:14,Choi:16,Hong:18,Jeon:TWC:18,Hong:CL:18,He:18,Studer:16}. For frequency flat channels, the optimal maximum-likelihood (ML) detection method and its low-complexity variations were developed in \cite{Wang:14,Choi:16,Hong:18,Jeon:TWC:18}. Particularly, in \cite{Hong:18,Jeon:TWC:18}, it was proven that the optimal ML detection is equivalent to the minimum weighted Hamming distance decoding in which the weights are determined by the likelihood functions. Utilizing this equivalence, the optimal soft-output detection method was proposed in \cite{Hong:CL:18} that computes a posteriori probability (APP) based on the weighted Hamming distance. For frequency selective channels, the optimal ML sequence detection method was developed by using Viterbi algorithm \cite{Hyowon:18} which is optimal in the sense of detecting the sequence of transmitted data symbols. Recently, the optimal soft-output detection method for frequency selective channels was proposed in \cite{Jeon:arXiv:18}, by utilizing the forward-backward algorithm based on a trellis diagram. In this work, a near-optimal low-complexity method was also developed based on the belief propagation algorithm by constructing a sparse factor graph. The common feature of the aforementioned methods is that they require the likelihood functions as the sufficient statistics for the data detection. Some suboptimal detection methods that can reduce the complexity of the optimal detection methods were presented in \cite{He:18,Studer:16}, but these methods are suboptimal and therefore suffer from a severe degradation in the detection performance. 

There is also a rich literature on channel estimation methods in MIMO systems with one-bit ADCs \cite{Choi:16,Studer:16,Li:2017,Lok:98,Mezghani:10,Stockle:16,Mo:18}. For frequency flat channels, an iterative channel estimation method was developed in \cite{Choi:16} on the basis of a ML criterion. In \cite{Li:2017}, a linear channel estimation method was developed  based on Bussgang's theorem \cite{Bussgang} which provides a linear representation of the quantized signals with Gaussian inputs. For frequency selective channels, iterative channel estimation algorithms were proposed based on the expectation-maximization algorithm \cite{Lok:98,Mezghani:10,Stockle:16} and the approximate-message-passing (AMP) algorithm \cite{Mo:18}. The common idea of these algorithms is to estimate unquantized signals and channel coefficients  separately and successively at each iteration. For quantized orthogonal frequency division multiplexing (OFDM) systems, a convex-optimization-based channel estimation algorithm was studied in \cite{Studer:16}, which provides the maximum a posteriori (MAP) estimate if the prior distribution of channel frequency responses is log concave. Despite all these efforts, obtaining an accurate channel information is still challenging in the MIMO systems with one-bit ADCs. The major reason is that coarse quantization at the ADCs fundamentally limits the available information at the receiver.

Recently, several different approaches beyond conventional data detection and channel estimation methods have studied for MIMO systems with one-bit ADCs \cite{Wen:16,Wang:17,Sun:18,Jeon:TVT:18}. In \cite{Wen:16}, a joint data-and-channel estimation technique was proposed on the basis of the bilinear generalized AMP (BiGAMP) algorithm to iteratively improve the estimation accuracy for both channel coefficients and data symbols, but its applicability is limited to frequency flat channels only. This limitation has overcome in \cite{Wang:17} by developing the Bayesian optimal data detector combined with a channel estimation method for MIMO OFDM systems with few-bit ADCs. The optimality of this method, however, is not guaranteed due to the use of OFDM signaling which is shown to be highly suboptimal when employing the few-bit ADCs \cite{Jeon:arXiv:18}. A joint channel estimation-and-decoding technique that does not rely on the OFDM signaling was proposed in \cite{Sun:18}. Unfortunately, this technique is still suboptimal in terms of the decoding performance, because it adopts the parametric BiGAMP algorithm based on Gaussian approximations. In \cite{Jeon:TVT:18}, inspired by the nonlinearity of the MIMO systems with few-bit ADCs, a supervised learning approach was proposed which learns the input-output relation of the nonlinear system by training examples and then uses the learned information for the data detection. One major limitation of this approach is that the length of the training sequence depends on the number of possible inputs; thereby, this approach may not be an efficient solution for the use in frequency selective channels.

\subsection{Contributions}
The major contributions of this paper are summarized as follows:
\begin{itemize}
    \item We present a likelihood function learning method for MIMO systems with one-bit ADCs. The key idea of the presented method is to exploit input-output samples obtained from the data detection, each describes the association between a quantized received vector and a transmitted symbol index at each time slot. Particularly, we define an empirical likelihood function that describes the empirical distribution of the input-output samples. We then exploit this empirical function to compensate a mismatch in a model-based likelihood function initially computed based on an estimated channel.  
    One prominent feature of the presented method is that it is universally applicable to any data detection method that utilizes the likelihood functions as the sufficient statistics, regardless of the channel estimation method, the frequency selectivity of the channel, and the type of the channel code adopted in the system.

    \item We optimize the presented learning method via a reinforcement learning approach, to resolve a label uncertainty problem in the samples caused by a data detection error. To this end, we formulate the optimization problem  as a Markov decision process (MDP) that maximizes the accuracy of the likelihood function learned from the input-output samples. Since the transition function of the MDP is unknown at the receiver, we develop a reinforcement learning algorithm that approximates the transition function and the optimal state of the MDP to find the optimal policy in a closed-form expression. The key advantage of the developed algorithm is that it is readily implemented in practical communication systems, unlike a conventional reinforcement learning algorithm. We also analyze the mean squared error (MSE) of the likelihood functions obtained from this policy. From the analysis results, we demonstrate that the mismatch in the likelihood function gradually reduces as the number of the input-output samples used for the learning increases.
    
    \item We also present two practical strategies to improve the performance of the presented method optimized by the reinforcement learning algorithm. The first strategy is to refine the input-output samples by reconstructing the transmitted symbol vectors at the receiver when cyclic redundancy check (CRC) bits are successfully decoded. Using this strategy, some \textit{false} input-output samples associating with symbol detection errors are refined into the \textit{true} samples that can be utilized to learn the likelihood function. 
    The second strategy is to generate \textit{virtual} input-output samples by exploiting the symmetric properties of the modulation alphabets and the noise distribution. Using this strategy, the number of the input-output samples is shown to increase  by four times for quadrature amplitude modulation (QAM) and the circularly symmetric noise (e.g., complex Gaussian noise). 
    
    \item Using simulations, we evaluate the performance gain achieved by using the proposed likelihood learning method for a coded MIMO system with one-bit ADCs under imperfect channel state information at the receiver (CSIR). In these simulations, the proposed method is applied to various data detection methods including the optimal ML detection  for frequency flat channels \cite{Wang:14,Choi:16}, and the optimal soft-output detection method and the low-complexity method for frequency selective channels  \cite{Jeon:arXiv:18}. Simulation results demonstrate that the proposed method significantly reduces the performance degradation caused by a mismatch in the likelihood function, regardless of the detection methods. One remarkable result is that the proposed method also provides a robustness to time-varying effects in wireless channels, by adapting the likelihood functions to channel variations. 
\end{itemize}

\subsubsection*{Notation}
Upper-case and lower-case boldface letters denote matrices and column vectors, respectively.
$\mathbb{E}[\cdot]$ is the statistical expectation,
$\mathbb{P}(\cdot)$ is the probability,
$(\cdot)^\top$ is the transpose,
$(\cdot)^{\sf H}$ is the conjugate transpose,
${\sf Re}\{\cdot\}$ is the real part,
${\sf Im}\{\cdot\}$ is the imaginary part,
$|\cdot|$ is the absolute value,
and $\Phi(\cdot)$ is the cumulative distribution of the standard normal random variable.
$({\bf a})_i$ represents the $i$-th element of a vector ${\bf a}$.
$\mathbb{I}\{\mathcal{A}\}$ is an indicator function which equals one if an event $\mathcal{A}$ is true and zero otherwise.
${\bf 0}_n$ is an $n$-dimensional vector whose elements are zero.

%

\begin{figure*}
	\centering 
	{\epsfig{file=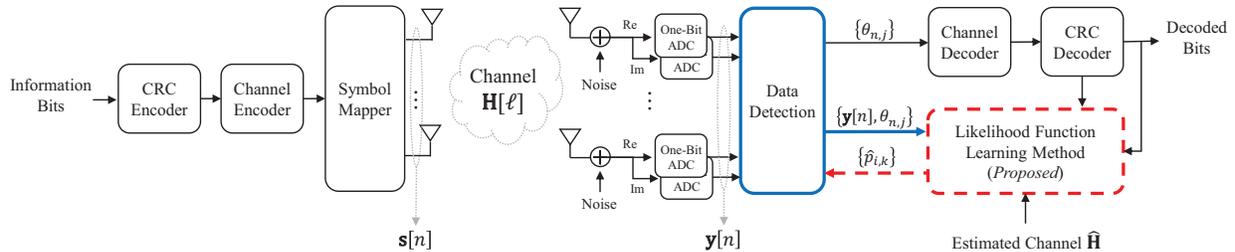, width=16.3cm}}
	\caption{A MIMO communication system with one-bit ADCs, in which a transmitter equipped with ${N}_{\rm tx}$ antennas communicates with a receiver equipped with $N_{\rm rx}$ antennas.} \vspace{-3mm}
	\label{fig:System}
\end{figure*}

\section{System Model and Preliminary}
In this section, we present a system model considered in this work. We then discuss the necessity and the challenge of likelihood function estimation to perform the optimal data detection in the considered system.

\subsection{System Model}
We consider a MIMO communication system with one-bit ADCs, in which a transmitter equipped with ${N}_{\rm tx}$ antennas communicates with a receiver equipped with $N_{\rm rx}$ antennas, as illustrated in Fig.~\ref{fig:System}. The wireless channel of the system is a frequency-selective channel described by an $L$-tap channel-impulse-responses (CIRs), where the number of CIR taps depends on the maximum delay spreads of the wireless channel and the signal bandwidth of the system. We denote the $l$-th CIR tap as ${\bf H}[\ell]\in\mathbb{C}^{N_{\rm rx}\times N_{\rm tx}}$ for $\ell \in\{0,1,\ldots,L-1\}$, where the $(i,j)$-th element of ${\bf H}[\ell]$ represents the $\ell$-th CIR\footnote{In mmWave communication systems, ${\bf H}[\ell]$ represents an \textit{effective} channel at the $\ell$-th discrete time delay, which abstracts the effects of an antenna array, transmit analog beamforming, and receive analog beamforming, as explained in \cite{Jeon:arXiv:18}.}  tap between the $i$-th receive antenna and the $j$-th transmit antenna. We assume a block-fading model in which each CIR tap keeps a constant value over a transmission frame, but in simulations, we also consider a time-varying channel model in which each CIR tap can change in a block.

\begin{figure*}
	\centering 
	{\epsfig{file=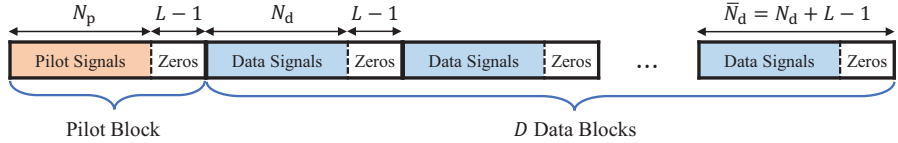, width=12cm}}
	\caption{A transmission frame that consists of one pilot signal block with length $N_{\rm p}$ and $D$ data blocks each with length $N_{\rm d}$ when $L-1$ zero vectors are appended at the end of every block.} \vspace{-3mm}
	\label{fig:Block}
\end{figure*}

We consider a transmission frame that consists of one pilot signal block with length $N_{\rm p}$ and $D$ data blocks each with length $N_{\rm d}$, as illustrated in Fig.~\ref{fig:Block}. To avoid inter-block-interference among different blocks, we assume that $L-1$ zero vectors are appended at the end of every block. In this work, we consider a time-domain signaling, because a frequency-domain signaling such as orthogonal-frequency-division multiplexing is suboptimal in one-bit ADC systems due to the nonlinearity of the quantization function at the ADCs, as discussed in \cite{Jeon:arXiv:18}. During the transmission of the pilot block, the transmitter sends pilot signals with length $N_{\rm p}$. Then the receiver uses the prior knowledge of the pilot signals to estimate $L$ CIR taps, $\{{\bf H}[\ell]\}_{\ell=1}^{L}$. During the transmission of each data block, the transmitter sends a sequence of a data symbol vector generated by successively applying 1) CRC appending, 2) channel encoding, and 3) symbol mapping to information bits. We denote the data symbol vector sent at time slot $n$ as ${\bf s}[n]\in \mathcal{X}^{N_{\rm tx}}$, where $\mathcal{X}$ is a constellation set. We assume that each data symbol vector satisfies a power constraint given by $\mathbb{E}[|({\bf s}[n])_i|^2]=1$ for $i\in\{1,\ldots,N_{\rm tx}\}$. Let ${\bar N}_{\rm d} \triangleq N_{\rm d}\!+\!L\!-\!1$ be the total duration of the received signal associating with each data block. Then the received signal at time slot $n$ before the ADCs is given by
\begin{align}\label{eq:unquant_sig}
	{\bf r}[n]
	&=\sum_{\ell=0}^{L-1}{\bf H}[\ell]{\bf s}[n-\ell] + {\bf z}[n] ={\bf H}{\bf x}[n] + {\bf z}[n],
\end{align}
where ${\bf H} = [{\bf H}[L-1], \cdots, {\bf H}[0]] \in \mathbb{C}^{N_{\rm rx}\times N_{\rm tx}L}$ is a concatenated channel matrix, ${\bf x}[n]$ is the effective symbol vector at time slot $n$, defined as ${\bf x}[n] = \big[{\bf s}^\top[n\!-\!L\!+\!1], \cdots, {\bf s}^\top[n]\big]^{\!\top}$,
and ${\bf z}[n]\sim \mathcal{CN}({\bf 0}_{N_{\rm rx}},{\sigma^2}{\bf I}_{N_{\rm rx}})$ is a circularly symmetric complex Gaussian noise vector at time slot $n$ with variance $\sigma^2$. Note that due to the zero padding at the end of each block, ${\bf s}[n]={\bf 0}_{N_{\rm tx}}$ for $n\notin \mathcal{N}_{d}$, where $\mathcal{N}_{d}\!=\!\{(d\!-\!1)\bar{N}_{\rm d}\!+\!1,\ldots,d\bar{N}_{\rm d}\}$,
which is the set of time slot indexes that associate with non-zero transmission at the $d$-th data symbol block for $d\in\{1,\ldots,D\}$. 
Using the effective symbol vector, we also define the $k$-th symbol vector, namely ${\bf x}_k$, as the $k$-th possible element in $\{{\bf x}[n]\!:\!\forall n\}$ for $k \in\mathcal{K}=\{1,\ldots,K\}$, where $K$ is the cardinality of $\{{\bf x}[n]\!:\!\forall n\}$.

At the ADCs, the real and imaginary parts of each element of the received signal in \eqref{eq:unquant_sig} are separately quantized using two independent one-bit scalar quantizers. Let ${\sf sign}: \mathbb{R}^{n} \rightarrow \{+1,-1\}^{n}$ be the quantization function of the scalar quantizer that maps an $n$-dimensional real-valued vector into an $n$-dimensional sign vector in $\{+1,-1\}^{n}$, where ${\sf sign}(r)=-1$ if $r\geq 0$ and ${\sf sign}(r)=1$ otherwise. Using this function, the quantized received vector at time slot $n$ is represented as
\begin{align}\label{eq:quant_sig}
	{\bf y}[n] = {\sf sign}\!\left({\sf Re}\big\{{\bf H}{\bf x}[n] + {\bf z}[n]\big\}\!\right) + j{\sf sign}\!\left({\sf Im}\big\{{\bf H}{\bf x}[n] + {\bf z}[n]\big\}\!\right).
\end{align}
The real-domain representation of the quantized vector is given by 
\begin{align}
\underbrace{\left[\!\!\begin{array}{c}
	{\sf Re}\{{\bf y}[n]\} \\ {\sf Im}\{{\bf y}[n]\} 
	\end{array}\!\!\right]}_{={\bf y}^{\sf re}[n]}
\!\!=\!{\sf sign}\Bigg(  \underbrace{\left[\!\!\begin{array}{cc}
	{\sf Re}\{{\bf H}\} \!\!&\!\! -{\sf Im}\{{\bf H}\} \\
	{\sf Im}\{{\bf H}\} \!\!&\!\! {\sf Re}\{{\bf H}\} \\
	\end{array}\!\!\right]}_{={\bf H}^{\sf re}}
\underbrace{\left[\!\!\begin{array}{c}
	{\sf Re}\{{\bf x}[n]\} \\ {\sf Im}\{{\bf x}[n]\} 
	\end{array}\!\!\right]}_{={\bf x}^{\sf re}[n]} + 
\underbrace{\left[\!\!\begin{array}{c}
	{\sf Re}\{{\bf z}[n]\} \\ {\sf Im}\{{\bf z}[n]\} 
	\end{array}\!\!\right]}_{={\bf z}^{\sf re}[n]} \Bigg).
\end{align}
To decode the $d$-th data block at the receiver, the set of the associating quantized vectors is used as an input of the data detection, which is given by $\{{\bf y}[n]\}_{n\in\mathcal{N}_d}$.

\subsection{Likelihood Function}
A likelihood function $\mathbb{P}\big({\bf y}[n]\big|{\bf x}[n]\!=\!{\bf x}_k\big)$ is the probability of receiving a quantized vector ${\bf y}[n]$ when assuming the $k$-th symbol vector ${\bf x}_k$ was sent at time slot $n$.
Perfect knowledge of the likelihood functions at the receiver is essential to realize a reliable communication in the MIMO system with one-bit ADCs, because these functions are the sufficient statistics of the optimal data detection methods \cite{Wang:14,Choi:16,Jeon:arXiv:18}. For example, in frequency flat channels ($L=1$), the ML estimate for the transmitted symbol vector at time slot $n$ is determined as 
\begin{align}\label{eq:MLD_k}
	\hat{\bf x}_{\sf ML}[n]= \argmax_{{\bf x}_k} ~\mathbb{P}\big({\bf y}[n]\big|{\bf x}[n]\!=\!{\bf x}_k\big).
\end{align}

In the MIMO system with one-bit ADCs specified in Section II-A, the likelihood function associating with the quantized vector ${\bf y}[n]$ and ${\bf x}_k$ is computed as 
\begin{align}\label{eq:LF_def}
	\mathbb{P}\big({\bf y}[n]\big|{\bf x}[n]\!=\!{\bf x}_k\big) 
	&= \prod_{i=1}^{2N_{\rm r}} \mathbb{P} \Big(  y_{i}^{\sf re}[n] 
	= {\sf sign}\big(({\bf h}_{i}^{\sf re})^{\!\top} {\bf x}_{k}^{\sf re} + z_{i}^{\sf re}[n]\big) \Big) \nonumber \\
	&= \prod_{i:y_{i}^{\sf re}[n]=+1} p_{k,i} \prod_{i:y_{i}^{\sf re}[n]=-1} (1-p_{k,i}),
\end{align}
where $({\bf h}_{i}^{\sf re})^{\!\top}$ is the $i$-th row of ${\bf H}^{\sf re}$, and $p_{i,k}$ is the element-wise likelihood function defined as
\begin{align}\label{eq:p_ik_def}
	p_{i,k}
	&= \mathbb{P} \big(y_{i}^{\sf re}[n]=+1 | {\bf x}[n]\!=\!{\bf x}_k\big) = \mathbb{P} \Big({\sf sign}\big((\bar{\bf h}_{i}^{\sf re})^{\!\top} {\bf x}_{k}^{\sf re} + z_{i}^{\sf re}[n]\big) \!=\! +1 \Big) = \Phi\left( \frac{({\bf h}_{i}^{\sf re})^{\!\top} {\bf x}_{k}^{\sf re}}{\sqrt{\sigma^2/2}} \right),
\end{align}
for $i\in\mathcal{I}=\{1,\ldots,2N_{\rm rx}\}$ and $k\in\mathcal{K}$.

Unfortunately, the perfect knowledge of the likelihood function at the receiver is not feasible in practical MIMO systems with one-bit ADCs due to imperfect CSIR. As can be seen in \eqref{eq:p_ik_def}, the likelihood functions are the function of the channel matrix, ${\bf H}$, but the receiver only knows an \textit{estimated} channel matrix, $\hat{\bf H}$, that contains an estimation error when using a pilot-assisted channel estimation method. Furthermore, this error is significant when employing the one-bit ADCs, because only the sign information of the received signal is available for the channel estimation at the receiver. The most common approach to deal with this problem is to simply ignore the channel estimation error and then to compute the likelihood functions based on the estimated channel. Then the likelihood function associating with the $k$-th symbol vector and the $i$-th quantized element is given by
\begin{align}\label{eq:p_mod}
	\hat{p}_{i,k}^{\sf mod} = \Phi\left( \frac{(\hat{\bf h}_{i}^{\sf re})^{\!\top} {\bf x}_{k}^{\sf re}}{\sqrt{\sigma^2/2}} \right),
\end{align}
where $({\bf h}_{i}^{\sf re})^{\!\top}$ is the $i$-th row of $\hat{\bf H}^{\sf re}$ defined as
\begin{align}\label{eq:H_tilded}
	\left[\begin{array}{cc}
	{\rm Re}\{{\bf \hat H}\} \!\!& -{\rm Im}\{{\bf \hat H}\} \\ {\rm Im}\{{\bf \hat H}\} \!\!& {\rm Re}\{{\bf \hat H}\} \\
	\end{array}\right],
\end{align}
for $i\in \mathcal{I}$ and $k\in\mathcal{K}$. In this work, we refer to the above estimate as a \textit{model-based} likelihood function, since it attempts to estimate the likelihood function based on the input-output model of the system. The optimal data detection methods using these model-based functions may suffer from a performance degradation due to a mismatch in the model caused by the channel estimation error.

\section{The Proposed Likelihood Function Learning Method}\label{Sec:Method}
In this section, we propose a likelihood function learning method that corrects mismatches in the model-based likelihood functions caused by a channel estimation error in MIMO systems with one-bit ADCs. 


\subsection{Basic Idea}\label{Sec:EmpLF} 
The basic idea of the proposed method is to update the model-based likelihood functions by exploiting input-output samples obtained from the data detection, each describes the association between the quantized received vector and the transmitted symbol index at each time slot. Our motivation is that the true likelihood function in \eqref{eq:p_ik_def} is represented by its empirical samples as follows:
\begin{align}\label{eq:p_limit}
	p_{i,k} 
	&= \frac{\mathbb{P}[ y_{i}^{\sf re}[n] =+1,{\bf x}[n]={\bf x}_k]}{\mathbb{P}[{\bf x}[n]={\bf x}_k]} \nonumber \\
	&\overset{(a)}{=} \lim_{\sum_{m\leq n} \mathbb{I}\{{k}[m]=k\} \rightarrow \infty} 
	\frac{\sum_{m \leq n}\mathbb{I}\{ y_{i}^{\sf re}[m] = +1,{k}[m]=k\}}{\sum_{m \leq n} \mathbb{I}\{{k}[m]=k\}} \nonumber \\
	&= \lim_{\sum_{m \leq n} \mathbb{I}\{{k}[m]=k\} \rightarrow \infty} 
		\frac{\sum_{m \leq n}\tilde{y}_i[m]\mathbb{I}\{{k}[m]=k\}}{\sum_{m \leq n} \mathbb{I}\{{k}[m]=k\}},
\end{align}
where (a) holds by the law of large numbers, $k[n]\in\mathcal{K}$ is the transmitted symbol index such that ${\bf x}[n]={\bf x}_{k[n]}$, and 
\begin{align}
	\tilde{y}_i[n] = \frac{1}{2}( y_{i}^{\sf re}[n] +1)
	= \begin{cases}
		1, & y_{i}^{\sf re}[n]=+1, \\
		0, & y_{i}^{\sf re}[n]=-1. \\	
	\end{cases}
\end{align}
Motivated from \eqref{eq:p_limit}, we define $(\tilde{\bf y}[n],\hat{k}[n])$ as the $n$-th input-output sample that describes the association between the quantized received vector and the transmitted symbol vector at time slot $n$, where $\tilde{\bf y}[n]=[\tilde{y}_1[n],\cdots, \tilde{y}_{2N_{\rm rx}}[n]]^\top \in\{0,1\}^{2N_{\rm rx}}$ and $\hat{k}[n]$ is the detected symbol index defined as
\begin{align}
	\hat{k}[n] = \argmax_{k\in\mathcal{K}}~ \theta_{n,k},
\end{align}
where $\theta_{n,k}$ is a-posteriori probability (APP) of the event $\{k[n]\!=\!k\}$ computed from the data detection based on the quantized observations at the receiver. We then define the empirical likelihood function associating with the $i$-th quantized element and the symbol index $k$ as 
\begin{align}\label{eq:p_emp}
	\hat{p}_{i,k}^{\sf emp} 
	= \frac{\sum_{m \leq n}(\tilde{\bf y}[n])_i\mathbb{I}\{\hat{k}[m]=k\}}{\sum_{m \leq n} \mathbb{I}\{\hat{k}[m]=k\}}  
	= \frac{u_{i,k}}{\bar{u}_{k}},
\end{align}
where 
\begin{align*}
	u_{i,k}= \sum_{m \leq n}(\tilde{\bf y}[n])_i\mathbb{I}\{\hat{k}[m]=k\},~~\text{and}~~\bar{u}_{k}= \sum_{m \leq n} \mathbb{I}\{\hat{k}[m]=k\}.
\end{align*}
A key advantage of the empirical likelihood function is that it approaches to the true function as the number of the input-output samples goes to the infinity, i.e., $\hat{p}_{i,k}^{\sf emp} \rightarrow p_{i,k}$ as $\bar{u}_{k} \rightarrow \infty$, provided that every symbol vector is correctly detected (i.e., $k[n]=\hat{k}[n]$, $\forall n$). To exploit this advantage, we propose a new estimate for the $(i,k)$-th likelihood function which is a linear combination of the empirical likelihood function in \eqref{eq:p_emp} and the model-based likelihood function in \eqref{eq:p_mod} as follows:
\begin{align}\label{eq:p_pro}
	\hat{p}_{i,k}^{\sf pro}(\alpha) & =  \alpha \hat{p}_{i,k}^{\sf mod} + (1-\alpha) \hat{p}_{i,k}^{\sf emp},
\end{align}
where $\alpha \in [0,1]$ is a combining ratio. The optimization of the combining ratio will be discussed in the sequel. 

A major factor that limits the accuracy of the proposed estimate in \eqref{eq:p_pro} is a \textit{label uncertainty} in the input-output samples which occurs when a data detection is incorrect;  
the detected symbol index in the sample may differ from the transmitted symbol index, i.e., $\hat{k}[n] \neq k[n]$. Under this uncertainty, the misusage of incorrect input-output samples may increase a mismatch in the empirical likelihood function. Therefore, a decision on the use of each input-output sample should be optimized to maximize the accuracy of the proposed estimate in \eqref{eq:p_pro}.

\subsection{Optimization Problem: Markov Decision Process}\label{Sec:MDP}
To overcome the limitation caused by the label uncertainty, we formulate an optimization problem that finds the optimal decision for each input-output sample to maximize the accuracy of the proposed estimate in \eqref{eq:p_pro}. Particularly, we formulate this problem as a Markov decision process (MDP) to capture the fact that the decision on the current input-output sample affects the decisions on the subsequent input-output samples. Each component of the MDP is defined below.

\subsubsection{State}
The state set of the MDP is defined as 
\begin{align}\label{eq:State_set}
	\mathcal{S}
	= \big\{ ({\bf U},{\bf W},n) ~\big|~& {\bf U}=[{\bf u}_1,\cdots,{\bf u}_K]\in \mathbb{Z}_+^{2N_{\rm rx}\times K}, 
	{\bf W}=[{\bf w}_1,\cdots,{\bf w}_K] \in \mathbb{Z}_+^{K\times K}, n \in \mathbb{Z}_+ \big\}.
\end{align}
where ${\bf w}_k=[w_{1,k},\cdots,w_{K,k}]^\top$ and
\begin{align}\label{eq:w_def}
	w_{j,k} = \sum_{m \leq n} \mathbb{I} \{k[m]=j, \hat{k}[m]=k\},
\end{align}
which represents the number of the quantized vectors associating with the transmitted symbol index $j$, but exploited to estimate the likelihood function associating with the symbol index $k$. Note that $\bar{u}_k =\sum_{m \leq n} \mathbb{I}\{\hat{k}[m]=k\} = \sum_{j=1}^K w_{j,k}$. 


\subsubsection{Actions}
The action set of the MDP is defined as
\begin{align}\label{eq:A_action_set}
	\mathcal{A} = \{ 0,1 \},
\end{align}
which indicates whether or not to update the proposed estimate by using the input-output sample associating with the current state. For example, if the action $a=1\in\mathcal{A}$ is associating with the state ${\bf S}=({\bf U},{\bf W},n) \in \mathcal{S}$, the $n$-th input-output sample $(\tilde{\bf y}[n],\hat{k}[n])$ is used to compute the empirical likelihood functions $\{\hat{p}_{i,\hat{k}[n]}^{\sf emp}\}_{i\in\mathcal{I}}$.

\subsubsection{Reward Function}
The reward function of the MDP for the states ${\bf S}, {\bf S}^\prime \in \mathcal{S}$ is defined as
\begin{align}\label{eq:Reward}
	{\sf R}({\bf S},{\bf S}^\prime) 
	= \sum_{i=1}^{2N_{\rm rx}}   \sum_{k=1}^K  \{{\sf MSE}_{i,k}^{\star}({\bf S}) - {\sf MSE}_{i,k}^{\star}({\bf S}^\prime) \},
\end{align}
where ${\sf MSE}_{i,k}^{\star}({\bf S}) = \min_{\alpha\in[0,1]} {\sf MSE}_{i,k}({\bf S};\alpha)$ with
\begin{align}\label{eq:MSE_def}
	{\sf MSE}_{i,k}({\bf S};\alpha) &= 
	\mathbb{E}_{\{{p}_{i,k}\}_k,\hat{p}_{i,k}^{\sf emp}}\!\Big[ \big| p_{i,k}-\hat{p}_{i,k}^{\sf pro}(\alpha) \big|^2 ~\Big|~{\bf w}_k \Big], 
\end{align}
provided that ${\bf S}=({\bf U},{\bf W},n)$. As can be seen in \eqref{eq:Reward}, the reward function ${\sf R}({\bf S},{\bf S}^\prime)$ is defined to quantify the improvement in the estimation error of the proposed estimate in terms of the MSE, when the state ${\bf S}$ is transited to the state ${\bf S}^\prime$. 

\begin{figure*}
	\centering 
	\subfigure[Original MDP]
	{\epsfig{file=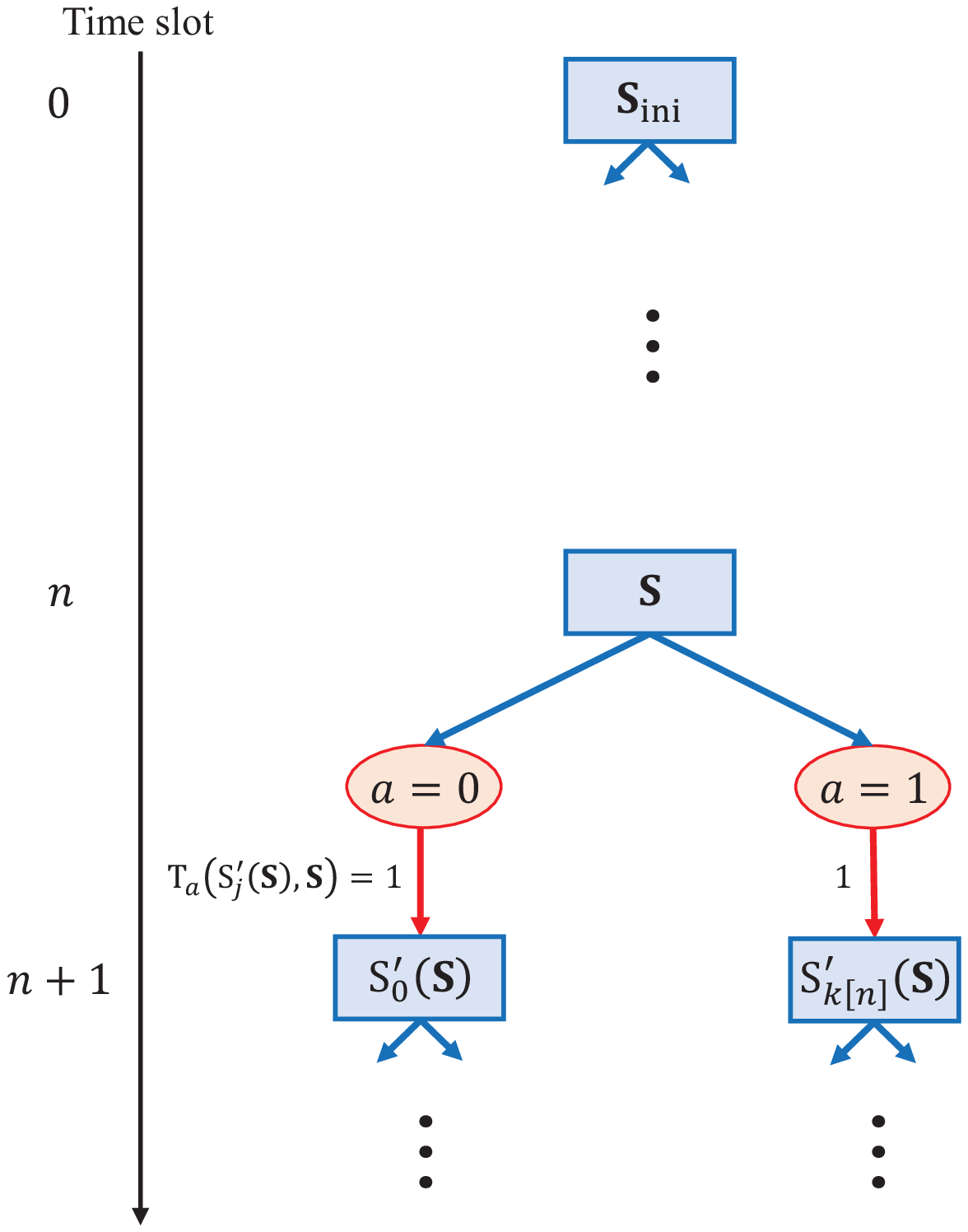, height=6cm}}
	\qquad\qquad\qquad
	\subfigure[Approximate MDP]
	{\epsfig{file=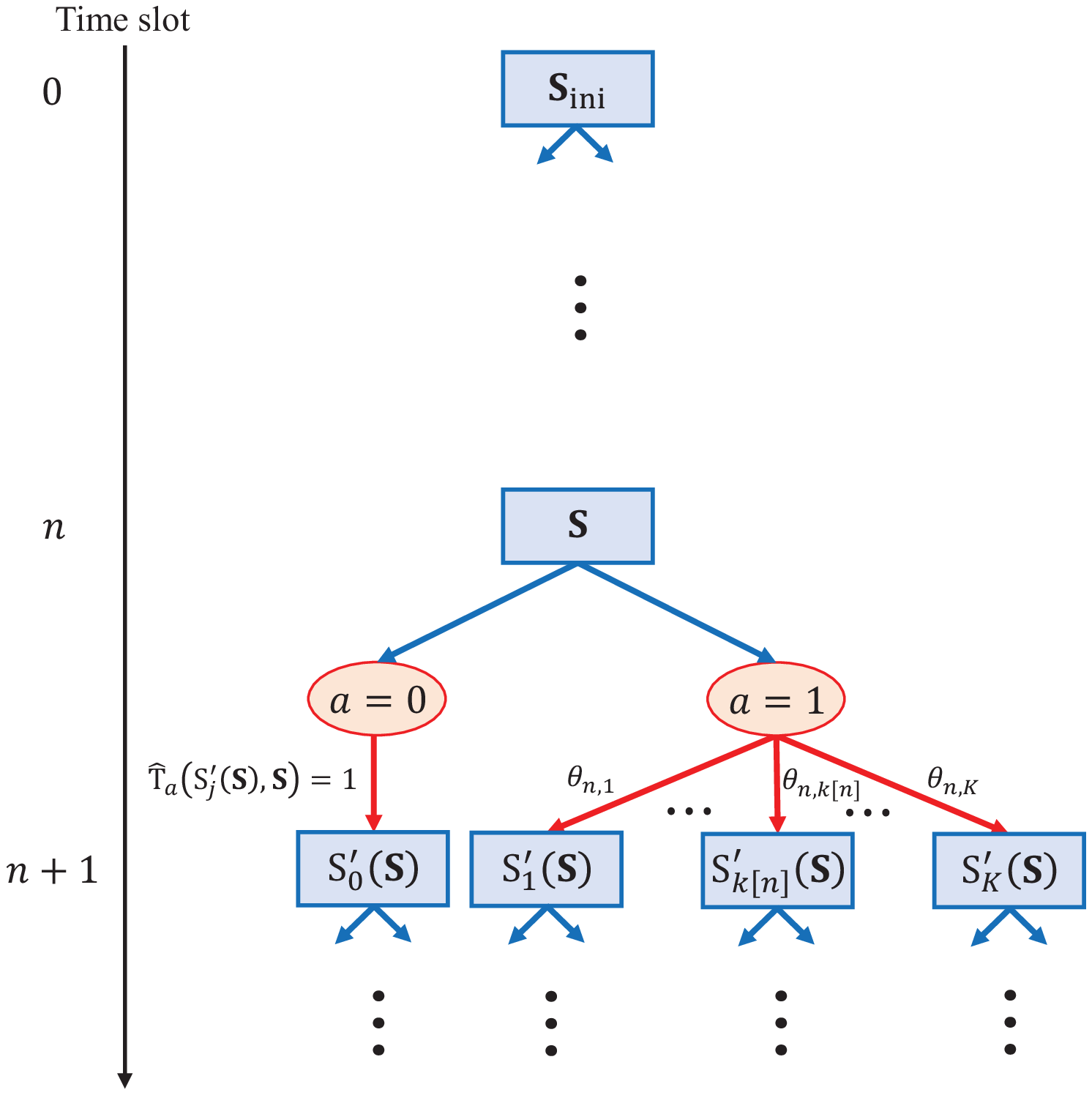, height=6cm}}
	\caption{The search tree of the MDP defined in Section IV-A with the original transition function in \eqref{eq:Trans2} (Fig.~\ref{fig:Tree}(a)) and with the approximate transition function in \eqref{eq:T_hat} (Fig.~\ref{fig:Tree}(b)).} \vspace{-3mm}
	\label{fig:Tree}
\end{figure*}

\subsubsection{Transition Function}
The (state) transition function of the MDP for the action $a\in\mathcal{A}$ and the states ${\bf S}, {\bf S}^\prime \in \mathcal{S}$ is defined as
\begin{align}\label{eq:Trans}
	{\sf T}_{a}({\bf S}^\prime,{\bf S}) 
	&= \mathbb{P}\big({\bf S}^{(n+1)} = {\bf S}^\prime \big| {\bf S}^{(n)}={\bf S},a\big), 
\end{align}
provided that ${\bf S}=({\bf U}, {\bf W},n)$ and ${\bf S}^\prime=({\bf U}^\prime, {\bf W}^\prime,n+1)$. If $a=0$, the $n$-th input-output sample will not be used to update the empirical likelihood functions, so it is obvious that ${\bf U}^\prime={\bf U}$ and ${\bf W}^\prime={\bf W}$. Whereas, if $a=1$, the $n$-th sample $(\tilde{\bf y}[n],\hat{k}[n])$ will be used to update the empirical likelihood functions $\{\hat{p}_{i,\hat{k}[n]}^{\sf emp}\}_{i\in\mathcal{I}}$. Therefore, by the definitions in \eqref{eq:p_emp} and \eqref{eq:w_def}, we have ${\bf U}^\prime={\bf U}+\tilde{\bf y}[n]{\bf e}_{\hat{k}[n]}^\top$ and ${\bf W}^\prime = {\bf W}+{\bf E}_{k[n],\hat{k}[n]}$, where ${\bf e}_k$ is the $k$-th column of ${\bf I}_K$ and ${\bf E}_{i,j}\in \{0,1\}^{K\times K}$ is a matrix with zero elements except its $(i,j)$-th position which is one. Based on these facts, the transition function in \eqref{eq:Trans} is rewritten as
\begin{align}\label{eq:Trans2}
	{\sf T}_{a}({\sf S}_{j}^\prime({\bf S}),{\bf S}) 
	&= \begin{cases}
	\mathbb{I}\{j=0\}, & a=0,\\
	\mathbb{I}\{k[n]=j\}, & a=1,\\
	\end{cases}
\end{align}
where ${\sf S}_{j}^\prime({\bf S})$ is the valid state that can be transited from ${\bf S}$, defined as
\begin{align}\label{eq:S_trans}
	{\sf S}_{j}^\prime({\bf S})
	& = \begin{cases}
	({\bf U},{\bf W},n+1), & j=0, \\
	({\bf U}+\tilde{\bf y}[n]{\bf e}_{\hat{k}[n]}^\top, {\bf W}+{\bf E}_{j,\hat{k}[n]},n+1), & j\in\mathcal{K}. \\
	\end{cases} 
\end{align}
The search tree of the MDP defined above is illustrated in Fig.~\ref{fig:Tree}(a).

The above MDP cannot be solved using dynamic programming in practical communication systems. The reason is that the transition function in \eqref{eq:Trans2} is unknown at the receiver due to the lack of the information of the transmitted symbol indexes. Furthermore, solving this MDP may require a prohibitive computational complexity because the number of the states exponentially increases with the number of the input-output samples, as can be seen in Fig.~\ref{fig:Tree}(a). Therefore, to solve the above MDP, it is essential to design a computationally-efficient algorithm that can perform without the perfect information of the transition function.

\subsection{Solving MDP: A Reinforcement Learning Approach}\label{Sec:RL}
Reinforcement learning is a well-known solution to solve an MDP with unknown transition and/or reward functions \cite{RL:Book}. Inspired by reinforcement learning, we present an optimization algorithm that approximately but efficiently solves the MDP defined in Section III-B. The key idea of the presented algorithm is to approximate both the transition function and the optimal states of the MDP to determine the optimal policy for each data block. A promising feature of this algorithm is that a mismatch in the optimal policy caused by the use of the approximation may gradually reduce as the algorithm is proceeded for multiple data blocks.



We first characterize the optimal policy for each data block in terms of the transition function and the optimal state of the MDP. Let ${\sf Q}({\bf S},a)$ be the Q-value associating with the state ${\bf S}\in\mathcal{S}$ and the action $a\in\mathcal{A}$. Then by the definition of the transition function in \eqref{eq:Trans2}, the Q-value ${\sf Q}({\bf S},a)$ is given by
\begin{align}\label{eq:Q_value}
	{\sf Q}({\bf S},a)
	&= \sum_{j\in\mathcal{K}\cup \{0\}} {\sf T}_{a}({\sf S}_{j}^\prime({\bf S}),{\bf S}) 
	\left\{ {\sf R}\big({\bf S},{\sf S}_{j}^\prime({\bf S})\big) + {\sf V}^\star\big({\sf S}_{j}^\prime({\bf S}) \big)\right\},
\end{align}
where ${\sf V}^{\star}({\bf S})$ is the sum of the future rewards when optimally acting from the state ${\bf S}$. Note that it is possible to assign more weight to the current reward than the future rewards by employing a discounting factor $\zeta \in [0,1]$. Nevertheless, in this work, we assume $\zeta = 1$ because every reward equally contributes to the improvement in the accuracy of the likelihood function estimate. Using the Q-value, the optimal policy for the state ${\bf S}\in\mathcal{S}$ is obtained as
\begin{align}\label{eq:Policy_opt0}
	\pi ({\bf S}) &= \argmax_{a \in \{0,1\} }~ {\sf Q}({\bf S},a).
\end{align}
From \eqref{eq:Q_value} and \eqref{eq:Policy_opt0}, the optimal policy for the state ${\bf S}$ is characterized as given in the following lemma:
\begin{lem}\label{thm:Policy_pro}
	The optimal policy for the state ${\bf S}\in\mathcal{S}$ associating with time slot $n\in \mathcal{N}_d$ is  
	\begin{align}\label{eq:Policy_opt}
		{\pi} ({\bf S}) &=\argmax_{a \in \{0,1\} } \sum_{i=1}^{2N_{\rm rx}} \sum_{j=0}^{K} 
		{\sf T}_{a}\big({\sf S}_{j}^\prime({\bf S}),{\bf S}\big) {\sf MSE}_{i,\hat{k}[n]}^{\star}\big({\sf S}_{j,d}^\star({\bf S})\big),
	\end{align}
	where ${\sf S}_{j,d}^\star({\bf S})$ is the optimal state when optimally acting from the state ${\sf S}_{j}^\prime({\bf S})$ for the samples from the $d$-th data block.
\end{lem}
\begin{IEEEproof}
	See Appendix A.
\end{IEEEproof}
\vspace{3mm}

Now, we derive a closed-form expression of the optimal policy in Lemma~1 by approximating the transition function and the optimal state. To approximate the transition function in \eqref{eq:Policy_opt},  we exploit the APP $\theta_{n,j}$ of the event $\{k[n]\!=\!j\}$ known at the receiver to estimate $\mathbb{I}\{k[n]\!=\!j\}$ in \eqref{eq:Trans2} unknown at the receiver. 
Using this strategy, we approximate the transition function ${\sf T}_{a}({\sf S}_{j}^\prime({\bf S}),{\bf S})$ in \eqref{eq:Trans2} as     
\begin{align}\label{eq:T_hat}
	\hat{\sf T}_{a}\big({\sf S}_{j}^\prime({\bf S}),{\bf S}\big) 
	= \begin{cases}
		\mathbb{I}\{j=0\}, & a=0,\\
		\theta_{n,j}, & a=1,\\
	\end{cases}
\end{align}
provided that ${\bf S}=({\bf U},{\bf W},n) \in\mathcal{S}$. The search tree of the MDP with this transition function is illustrated in Fig.~\ref{fig:Tree}(b). Based on the approximate transition function in \eqref{eq:T_hat}, it is possible to determine the optimal state ${\sf S}_{j,d}^\star({\bf S})$ in \eqref{eq:Policy_opt} via the dynamic programming, as done in conventional model-based reinforcement learning \cite{RL:Book}. This approach, however, may require a prohibitive computational complexity because the number of states in the MDP exponentially increases with a data block length, as can be seen from Fig.~\ref{fig:Tree}. To develop a practical algorithm that can be implemented in communication systems, we also approximate the optimal state by considering the ideal case in which all the symbol indexes after time slot $n\!+\!1$ are correctly detected, i.e., $k[m]=\hat{k}[m]$ for $n+1\leq m\in \mathcal{N}_d$. In this case, the optimal state ${\sf S}_{j,d}^\star({\bf S})$ is readily computed as
\begin{align}\label{eq:S_opt_hat}
	\hat{\sf S}_{j,d}^\star({\bf S})
	& = \begin{cases}
	({\bf U}^\star, {\bf W}^\star,d\bar{N}_{\rm d}\!+\!1), &\!\! j=0, \\
	({\bf U}^\star+\tilde{\bf y}[n]{\bf e}_{\hat{k}[n]}^\top, {\bf W}^\star+{\bf E}_{j,\hat{k}[n]},d\bar{N}_{\rm d}\!+\!1), &\!\! j\in\mathcal{K}, \\
	\end{cases} 
\end{align}
where 
	${\bf U}^\star = {\bf U} \!+\! \sum_{m =n+1}^{d\bar{N}_{\rm d}}  \tilde{\bf y}[m]{\bf e}_{\hat{k}[m]}^\top$,
	${\bf W}^\star = {\bf W} \!+\! \sum_{k\in\mathcal{K}}{\delta}_{n,k}  {\bf E}_{k,k}$, and
	${\delta}_{n,k} = \sum_{m = n+1}^{d\bar{N}_{\rm d}} \mathbb{I}\big\{\hat{k}[m]=k\big\}$.
Utilizing the above fact, we approximate the optimal state ${\sf S}_{j,d}^\star({\bf S})$ as $\hat{\sf S}_{j,d}^\star({\bf S})$. From the approximations in \eqref{eq:T_hat} and \eqref{eq:S_opt_hat}, we obtain a closed-form expression of the optimal policy as given in the following theorem:
\begin{thm}\label{lem:MSE_pro}
	If ${\sf T}_{a}({\sf S}_{j}^\prime({\bf S}),{\bf S})$ and ${\sf S}_{j,d}^\star({\bf S})$ are given as in \eqref{eq:T_hat} and \eqref{eq:S_opt_hat}, respectively,     
	the optimal policy for the state ${\bf S}\in\mathcal{S}$ associating with time slot $n \in\mathcal{N}_d$ is  
	\begin{align}\label{eq:Policy_pro}
		{\pi} ({\bf S})
		&= \mathbb{I}\left[  \sum_{i=1}^{2N_{\rm rx}}(\hat{\mathcal{E}}_{i,k}^{\sf mod})^2 
			\!\left\{\sum_{j=1}^{K} \theta_{n,j}
			\frac{\big(w_{k,k}^\star \!+\!\mathbb{I}\{j\!=\!k\}\big)^2} {\Omega_{i,k}^{({\bf w}_k^\star + {\bf e}_j)}} 
			- \frac{(w_{k,k}^\star)^2 } {\Omega_{i,k}^{({\bf w}_k^\star)}} \right\}\right]\!, 
	\end{align}
	where $k=\hat{k}[n]$, $w_{k,k}^\star=w_{k,k}\!+\!{\delta}_{n,k}$, ${\bf w}_{k}^\star={\bf w}_k\!+\!{\delta}_{n,k}{\bf e}_k$,
	\begin{align*}	
		\Omega_{i,k}^{({\bf w}_k)} &= \bigg(\sum_{j\neq k} w_{j,k}\hat{\Delta}_{i,k,j}^{\sf mod} \bigg)^2 
			+ \sum_{j} w_{j,k} \hat{v}_{i,j} + \sum_{j} w_{j,k}^2\hat{\mathcal{E}}_{i,j}^{\sf mod}, \nonumber \\
		\hat{\Delta}_{i,k,j}^{\sf mod} &= \hat{p}_{i,k}^{\sf mod}-\hat{p}_{i,j}^{\sf mod}, \nonumber \\
		\hat{v}_{i,j} &= \hat{p}_{i,j}^{\sf mod}(1\!-\!\hat{p}_{i,j}^{\sf mod}) - \hat{\mathcal{E}}_{i,j}^{\sf mod},	
	\end{align*}
	provided that ${\bf S}=({\bf U},{\bf W},n)$, $\mathbb{E}[p_{i,k}]=\hat{p}_{i,k}^{\sf mod}$, and ${\sf Var}(p_{i,k})=\hat{\mathcal{E}}_{i,k}^{\sf mod}$ for $i\in\mathcal{I}$ and $k\in\mathcal{K}$. 
\end{thm}
\begin{IEEEproof}
	See Appendix B.
\end{IEEEproof}
\vspace{3mm}
The common feature of the approximations adopted in Theorem~1 is that their tightness increases as the data detection performance improves because $\hat{\sf T}_{a}({\sf S}_{j}^\prime({\bf S}),{\bf S}) \rightarrow {\sf T}_{a}({\sf S}_{j}^\prime({\bf S}),{\bf S})$ and $\hat{\sf S}_{j,d}^\star({\bf S}) \rightarrow {\sf S}_{j,d}^\star({\bf S})$ as $\theta_{n,k[n]} \rightarrow 1$ for $n\in\mathcal{N}_d$. This implies that a mismatch in the policy caused by the use of the approximations in \eqref{eq:T_hat} and \eqref{eq:S_opt_hat} can be reduced by improving the accuracy of the likelihood function estimates. Fortunately, the accuracy of the proposed estimates is expected to increase as the number of the input-output samples increases. Therefore, the optimal policy in Theorem~1 becomes close to  the \textit{true} optimal policy as the presented algorithm is proceeded for multiple data blocks within the channel coherence time.


After the detection of the $d$-th data block, the receiver updates the current (or initial) state according to the optimal policy in Theorem~1 for a set of the input-output samples $\{(\tilde{\bf y}[n],\hat{k}[n])\}_{n\in\mathcal{N}_d}$ and the corresponding APPs $\{\theta_{n,j}\}_{n\in\mathcal{N}_d,j\in\mathcal{K}}$ obtained from the data detection. Suppose that the state is given by ${\bf S}\in\mathcal{S}$ after the update for the data block $d$. Then by using the result in \eqref{eq:Apdx:alpha_compute}, the proposed likelihood function in \eqref{eq:p_pro} is determined as
\begin{align}\label{eq:p_pro2}
	\hat{p}_{i,k}^{\sf pro}({\bf S}) 
	& =  \left(1 - \frac{ w_{k,k}\bar{u}_k \hat{\mathcal{E}}_{i,k}^{\sf mod}  }  {   \Omega_{i,k}^{({\bf w}_k)}  }\right) \hat{p}_{i,k}^{\sf mod} 
	  + \frac{ w_{k,k}\hat{\mathcal{E}}_{i,k}^{\sf mod}  }  {   \Omega_{i,k}^{({\bf w}_k)}  } u_{i,k},
\end{align}
where $\bar{u}_k = \sum_{j}w_{j,k}$ provided that ${\bf S} = ({\bf U},{\bf W},d\bar{N}_d\!+\!1)$.

\begin{algorithm}
	\caption{The proposed likelihood function learning method.}\label{alg:Sum}
	{\small{\begin{algorithmic}[1]
				\STATE Initialize $p_{i,k}=\hat{p}_{i,k}^{\sf mod}$ from \eqref{eq:p_mod} based on $\hat{\bf H}$, $\forall i\in\mathcal{I},k\in\mathcal{K}$.
				\STATE Compute $\hat{\mathcal{E}}_{i,k}^{\sf mod}$ using an offline learning process based on $\hat{p}_{i,k}^{\sf mod}$ and $\hat{\bf H}$, $\forall i\in\mathcal{I},k\in\mathcal{K}$. 
				\STATE Initialize ${\bf S}=\big( {\bf 0}_{2N_{\rm rx}\times K}, {\bf 0}_{K\times K}, 1\big)$.
				\FOR {$d=1$ to $D$}	
				\STATE Perform a data detection based on $\{p_{i,k}\}_{i,k}$. 
				\STATE Set $\{(\tilde{\bf y}[n],\hat{k}[n])\}_{n\in\mathcal{N}_d}$ and $\{\theta_{n,j}\}_{n\in\mathcal{N}_d,j\in\mathcal{K}}$. 
				\FOR {$n\in \mathcal{N}_{d}$}
				\STATE Compute ${\pi}({\bf S})$ from \eqref{eq:Policy_pro} with $\hat{v}_{i,k}=\hat{p}_{i,k}^{\sf mod}(1-\hat{p}_{i,k}^{\sf mod})$, $\forall i\in\mathcal{I},k\in\mathcal{K}$. 
				\STATE Update ${\bf S} \leftarrow {\sf S}_{\hat{k}[n]}^\prime({\bf S})$ if ${\pi}({\bf S})=1$, 
				and ${\bf S} \leftarrow {\sf S}_{0}^\prime({\bf S})$ if ${\pi}({\bf S})=0$ from \eqref{eq:Trans2}. 
				\ENDFOR
				\STATE Update $p_{i,k} \leftarrow \hat{p}_{i,k}^{\sf pro}({\bf S})$ from \eqref{eq:p_pro2} with $\hat{v}_{i,k}=\hat{p}_{i,k}^{\sf mod}(1-\hat{p}_{i,k}^{\sf mod})$, $\forall i\in\mathcal{I},k\in\mathcal{K}$.
				\ENDFOR
	\end{algorithmic}}}
\end{algorithm}

In Algorithm~1, we summarize the proposed likelihood function learning method optimized via the presented algorithm. In Step 2, an offline learning process is adopted to compute the MSE of the model-based likelihood function, which will be discussed with more details in Section~\ref{Sec:Offline}. In Step 9, the current state ${\bf S}$ is updated according to the optimal action determined from \eqref{eq:Policy_pro}. Particularly, since the knowledge of the next state is neither observable nor available at the receiver, if ${\pi}({\bf S})=1$, the most probable transition is assumed, i.e., ${\bf S} \leftarrow {\sf S}_{\hat{k}[n]}^\prime({\bf S})$. In Step~8 and Step~11 of Algorithm~1, we approximate $\hat{v}_{i,k} = \mathbb{E}[{p}_{i,k}(1-{p}_{i,k})]$ as $\hat{p}_{i,k}^{\sf mod}(1-\hat{p}_{i,k}^{\sf mod})$ to guarantee that $\hat{v}_{i,k}\geq 0$ which may not hold under our assumptions of $\mathbb{E}[{p}_{i,k}]=\hat{p}_{i,k}^{\sf mod}$ and ${\sf Var}({p}_{i,k})=\hat{\mathcal{E}}_{i,k}^{\sf mod}$. 


\vspace{2mm}
{\bf Remark (Comparison to Model-Based Reinforcement Learning):} 
The presented algorithm resembles to a conventional model-based reinforcement learning algorithm: 1) both algorithms attempt to approximate an unknown MDP first and then find the optimal policy based on the approximate MDP; and 2) these two steps are repeated to improve the policy, where in the presented algorithm the decision for each data block corresponds to each iteration. Despite this resemblance, our algorithm also has some key differences. In the conventional algorithm, the MDP is empirically learned by a training process, and the (approximate) optimal policy is obtained via the dynamic programming. Whereas, in our algorithm, the MDP is approximately learned by the APPs obtained from the data detection, and the (approximate) optimal policy is derived in a closed-form expression. These differences are essential to reduce the computational complexity of the presented algorithm, so that it can be readily implemented in practical communication systems. Therefore, the presented algorithm can be regarded as a low-complexity variation of the model-based reinforcement learning for the application in communication systems.

\subsection{Mean-Squared-Error (MSE) Analysis}\label{Sec:RL:MSE}
We also analyze the reduction in the estimation error achieved when using the proposed method. 
The result is given in the following corollary:
\begin{cor}\label{cor:Analysis}		
	As $\beta_k \triangleq \frac{w_{k,k}}{\sum_{j}w_{j,k}}\rightarrow 1$, the MSE defined in \eqref{eq:MSE_def} becomes
	\begin{align}\label{eq:MSE_limit}
		\lim_{\beta_k  \rightarrow 1} {\sf MSE}_{i,k}^\star({\bf S}) 
		=\frac{  \hat{v}_{i,k}\hat{\mathcal{E}}_{i,k}^{\sf mod} }	
			{  \hat{v}_{i,k} + w_{k,k} \hat{\mathcal{E}}_{i,k}^{\sf mod} } ,
	\end{align}
	provided that ${\bf S}=({\bf U},{\bf W},n)$, $\mathbb{E}[p_{i,k}]=\hat{p}_{i,k}^{\sf mod}$, and ${\sf Var}(p_{i,k})=\hat{\mathcal{E}}_{i,k}^{\sf mod}$.
\end{cor}
\begin{IEEEproof}
	Since $\beta_k = \frac{w_{k,k}}{\sum_{j}w_{j,k}}\rightarrow 1$ implies that $w_{j,k}\rightarrow 0$ for $j\neq k \in\mathcal{K}$, the above result is directly obtained from \eqref{eq:Apdx:MSE_compute} derived in Appendix B. 
\end{IEEEproof}
\vspace{3mm}
\noindent Corollary~\ref{cor:Analysis} shows that the MSE of the likelihood function obtained by the proposed method decreases with the number of the exploited input-output samples. Particularly, if $\hat{\mathcal{E}}_{i,k}^{\sf mod}>0$, this error approaches to zero as the number of the input-output samples goes to the infinity. Therefore, the analysis result demonstrates that the proposed method has a potential to realize the perfect knowledge of the likelihood functions at the receiver even in the imperfect CSIR case. Although this advantage is attained only when the channel coherence time is sufficiently long, we also demonstrate that the proposed method is beneficial even in time varying channels, as will be shown in Section~\ref{Sec:Simul}. 

\section{Performance Improvement Stratigies}\label{Sec:Improve}
In this section, we present two practical strategies to improve the effectiveness of the proposed likelihood function learning method in Section~\ref{Sec:Method}. We also introduce a simple offline learning method to learn the MSE of the model-based likelihood function, required by the proposed method. 

\subsection{Sample Refinement Using CRC}
We present a sample refinement strategy that exploits CRC bits to refine incorrect input-output samples (i.e., the samples with $\hat{k}[n]\neq k[n]$) which limit the performance of the proposed likelihood function learning method. The key idea is to reconstruct the transmitted symbol vectors by applying the transmission procedures to the decoded bits at the receiver, only when the CRC bits are successfully decoded. Suppose that the CRC bits appended to the $d$-th data block are successfully decoded. Using the above strategy, for $n\in\mathcal{N}_d$, the $n$-th input-output sample $(\tilde{\bf y}[n],\hat{k}[n])$ is refined into $(\tilde{\bf y}[n],{k}_{\rm rec}[n])$ such that ${\bf x}_{\rm rec}[n]={\bf x}_{k_{\rm rec}[n]}$, where ${\bf x}_{\rm rec}[n]$ is the $n$-th reconstructed symbol vector at the receiver. The corresponding APP $\theta_{n,j}$ is also refined into
\begin{align}\label{eq:refine_post}
	\theta_{n,j} \!=\! 
	\begin{cases}
	1, & j={k}_{\rm rec}[n], \\ 
	0, & j \neq k_{\rm rec}[n],
	\end{cases}
\end{align}
for $n\in\mathcal{N}_d$. If the CRC bits of the $d$-th data block are sufficient to check any error in the decoded bits, the reconstructed symbol vectors are the same as the transmitted symbol vectors (i.e.,  $k_{\rm rec}[n]=k[n]$, $\forall n\in\mathcal{N}_d$). In this case, all the input-output samples associating with the $d$-th data block become the true input-output samples. Therefore, by applying the presented strategy, the proposed learning method can utilize more number of the input-output samples. In addition, this strategy also improves the tightness of the approximation adopted in the reinforcement learning algorithm in Section~\ref{Sec:RL}, because the use of the refined APPs reduces the mismatch in the approximate transition function in \eqref{eq:T_hat}.

\subsection{Virtual-Sample Generation Using Symmetric Property}
We also present a virtual-sample generation strategy that generates additional input-output samples by exploiting the symmetric properties of the modulation alphabets and the noise distribution.
The required symmetric properties are as follows: 1) the modulation alphabets should be symmetric with respect to the origin, the in-phase axis, and the quadrature axis in the constellation diagram (e.g., QAM), and also 2) the distribution of the noise should be circularly symmetric (e.g., $\mathcal{CN}(0,\sigma^2)$).
If the modulation set $\mathcal{X}$ holds the first condition, for any $M$-dimensional symbol vector ${\bf x}_k\in \mathcal{X}^M$, there also exist three symbol vectors $-{\bf x}_k,j{\bf x}_k,-j{\bf x}_k \in \mathcal{X}^M$. Then by the circularly symmetric property of the noise distribution, the following equalities are obtained:
\begin{align}\label{eq:symmetry}
	\mathbb{P}({\bf y}[n]|{\bf x}[n]\!=\!{\bf x}_k) 
	= \mathbb{P}(-{\bf y}[n]|{\bf x}[n]\!=\!-\!{\bf x}_{k})  
	= \mathbb{P}(j{\bf y}[n]|{\bf x}[n]\!=\!j{\bf x}_{k}) 
	= \mathbb{P}(-j{\bf y}[n]|{\bf x}[n]\!=\!-\!j{\bf x}_{k}),
\end{align}
as shown in \cite{Jeon:TVT:18}. The above equalities imply that the $n$-th input-output sample $(\tilde{\bf y}[n],\hat{k}[n])$ can be used to generate three virtual samples: $(-\tilde{\bf y}[n],\hat{k}_1[n])$, $(\tilde{\bf y}^*[n],\hat{k}_2[n])$, and $(-\tilde{\bf y}^*[n],\hat{k}_3[n])$, where 
\begin{align}\label{eq:y_conj}
	\tilde{\bf y}^*[n] = 
	\left[\!\!\begin{array}{c}
	(\tilde{\bf y}[n])_{N_{\rm rx}+1:2N_{\rm rx}} \\ -(\tilde{\bf y}[n])_{1:N_{\rm rx}}
	\end{array}\!\!\right],
\end{align}
and $\hat{k}_1[n],\hat{k}_2[n],\hat{k}_3[n] \in \mathcal{K}$ are the symbol indexes such that ${\bf x}_{\hat{k}_1[n]}\!=\!-{\bf x}_{\hat{k}[n]}$, ${\bf x}_{\hat{k}_2[n]}\!=\!j{\bf x}_{\hat{k}[n]}$, ${\bf x}_{\hat{k}_3[n]}\!=\!-j{\bf x}_{\hat{k}[n]}$. Therefore, every time the $n$-th sample $(\tilde{\bf y}[n],\hat{k}[n])$ is utilized to update the empirical likelihood function in the proposed method, three virtual samples $(-\tilde{\bf y}[n],\hat{k}_1[n])$, $(\tilde{\bf y}^*[n],\hat{k}_2[n])$, and $(-\tilde{\bf y}^*[n],\hat{k}_3[n])$ can also be utilized, which improve the sample efficiency of the proposed method by four times.

\subsection{Offline Learning for Initial Estimation Error}\label{Sec:Offline}
We present a simple offline learning method to learn the MSE of the model-based likelihood function in \eqref{eq:p_mod}, which is necessary to use the reinforcement learning algorithm in Section~\ref{Sec:RL}. The basic idea is to generate multiple \textit{pseudo} channels by applying a channel estimation method to the initial estimated channel. Then the MSEs are learned by averaging the squared errors between the model-based likelihood functions computed based on the pseudo channels and the initial channel.

Let ${\bf X}_{\rm p}=\big[{\bf x}_{\rm p}[1],\ldots,{\bf x}_{\rm p}[N_{\rm p}]\big] \in\mathbb{C}^{N_{\rm tx}\times N_{\rm p}}$ be a pilot signal matrix used in the channel estimation, where ${\bf x}_{\rm p}[n] \in \mathbb{C}^{N_{\rm tx}}$ is the $n$-th pilot signal vector, and $N_{\rm p}$ is the length of the pilot signals such that $N_{\rm p}\geq L(N_{\rm tx}\!-\!1)\!+\!1$. By regarding $\hat{\bf H}$ as a true channel matrix, the receiver generates $T_{\rm train}$ pseudo quantized matrices; each corresponds to a quantized received matrix obtained when transmitting the pilot signals through the channel $\hat{\bf H}$. The $t$-th pseudo quantized matrix is given by 
\begin{align}\label{eq:CE_received}
	{\bf Y}_{\rm p}^{(t)}  = {\sf sign}\big({\rm Re}\{\bar{\bf X}_{\rm p} \hat{\bf H}^{\sf H} + {\bf Z}_{\rm p}^{(t)}\}\big) 
	+ j{\sf sign}\big({\rm Im}\{\bar{\bf X}_{\rm p} \hat{\bf H}^{\sf H} + {\bf Z}_{\rm p}^{(t)}\}\big),
\end{align}
where $\bar{\bf X}_{\rm p} \in \mathbb{C}^{(N_{\rm p}+L-1)\times LN_{\rm tx}}$ is a Toeplitz-type matrix that consists of the pilot signals, and ${\bf Z}_{\rm p}^{(t)}\in \mathbb{C}^{(N_{\rm p}+L-1)\times N_{\rm rx}}$ is the $t$-th pseudo noise matrix generated according to the noise distribution. Then the $t$-th pseudo channel matrix, namely $\hat{\bf H}^{(t)}\in\mathbb{C}^{N_{\rm rx}\times N_{\rm tx}L}$, is obtained by applying the channel estimation method to ${\bf Y}_{\rm p}^{(t)}$, for $t\in\{1,\ldots,T_{\rm train}\}$. Using these $T_{\rm train}$ channel matrices, we estimate the MSE  of the model-based likelihood function as follows:
\begin{align}\label{eq:learn_mod}
	\hat{\mathcal{E}}_{i,k}^{\sf mod} 
	&\approx \frac{1}{T_{\rm train}}\sum_{t=1}^{T_{\rm train}}\left(  \Phi\left( \frac{(\hat{\bf h}_{{\rm R},i}^{(t)})^\top {\bf x}_{k}^{\sf re}}{\sqrt{\sigma^2/2}} \right) - \hat{p}_{i,k}^{\sf mod} \right)^2,
\end{align}
for $i\in\{1,\ldots,2N_{\rm rx}\}$ and $k\in\mathcal{K}$, where $(\hat{\bf h}_{{\rm R},i}^{(t)})^\top$ is the $i$-th row of 
\begin{align}
	\left[\begin{array}{cc}
		{\rm Re}\{\hat{\bf H}^{(t)}\} \!\!& -{\rm Im}\{\hat{\bf H}^{(t)}\} \\ {\rm Im}\{\hat{\bf H}^{(t)}\} \!\!& {\rm Re}\{\hat{\bf H}^{(t)}\} \\
	\end{array}\right].
\end{align}

\section{Simulation Results}\label{Sec:Simul}
In this section, using simulations, we evaluate the performance gain achieved by the proposed likelihood function learning method when it is applied to various data detection methods in a MIMO system with one-bit ADCs. In these simulations, we adopt 4-QAM for the symbol mapping, and 16-bit CRC bits with the polynomial of $z^{16} + z^{15} + z^2 + 1$. For channel coding, we adopt the rate $\frac{1}{2}$ turbo codes based on parallel concatenated codes with feedforward and feedback polynomial (15,13) in octal notation. For the proposed method, we apply the performance improvement strategies and the offline learning process, both presented in Section~\ref{Sec:Improve}, with $T_{\rm train}=10$. In Step 2 of Algorithm~1, we set $\hat{\mathcal{E}}_{i,k}^{\sf mod}=\max\{\hat{\mathcal{E}}_{i,k}^{\sf mod},10^{-20}\}$, $\forall i\in\mathcal{I},k\in\mathcal{K}$ to improve numerical stability.

\subsection{Frequency-Flat Channels}
We present simulation results for Rayleigh-fading frequency-flat channels. In this simulation, we consider three data detection methods: 1) the maximum-likelihood (ML) detection \cite{Wang:14,Choi:16}, 2) the GAMP-based detection\footnote{In this method, we perform a joint channel-and-data estimation algorithm in \cite{Wen:16} for the given estimated channel matrix, by setting all signals as data signals (i.e., $N_{\rm p}=0$).} \cite{Wen:16}, and 3) zero-forcing (ZF) detection. We refer to the ML detection operating with the proposed likelihood learning method as \textit{robust ML}. For the channel estimation, we adopt a linear minimum-MSE (LMMSE) method with $N_{\rm p}$ pilot signals which ignores the quantization effect at the ADCs. 

\begin{figure}
	\centering 
	{\epsfig{file=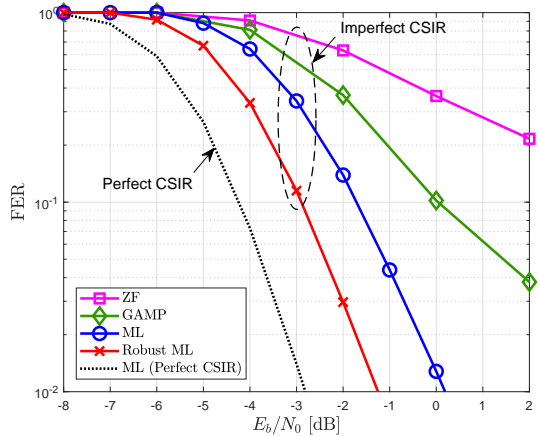, width=8cm}}
	\caption{The FER vs. $E_b/N_0$ of the ML, the robust ML, the GAMP-based, and the ZF detection methods for a time-invariant frequency-flat channel in a $4\times 8$ MIMO system with one-bit ADCs.} \vspace{-3mm}
	\label{fig:FER_4x8_Flat}
\end{figure}

Fig.~\ref{fig:FER_4x8_Flat} compares the FERs of various detection methods for a time-invariant frequency-flat channel in a $4\times 8$ MIMO system with one-bit ADCs. The parameters related to the transmission frame are set to be $N_{\rm p}=8N_{\rm tx}$, $D=40$, and $N_{\rm d}=128$ (i.e., 1024 coded bits). As can be seen in Fig.~\ref{fig:FER_4x8_Flat}, the ML detection with perfect CSIR achieves the optimal FER, but this method suffers from a severe performance loss under imperfect CSIR due to the mismatch in the model-based likelihood function. Whereas, the robust ML detection effectively reduces this loss by applying the proposed method to correct the likelihood function mismatch; thereby, in the imperfect CSIR case, the robust ML detection achieves the lowest FER among all the considered detection methods. Other conventional detection methods (GAMP-based detection and ZF detection) are not only suboptimal in terms of the FER performance, but also vulnerable to the effect of imperfect CSIR. Therefore, these methods are inferior to both the robust and the conventional ML detection methods.

\begin{figure}
	\centering 
	{\epsfig{file=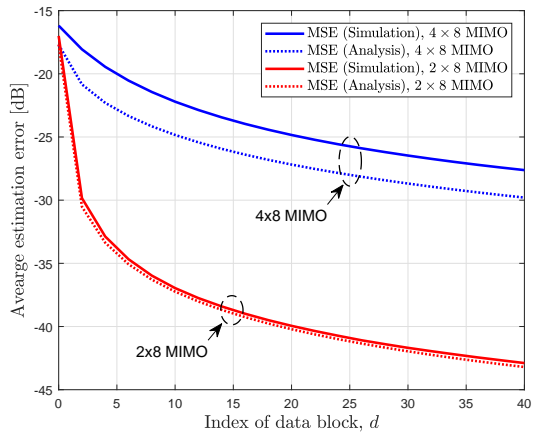, width=8cm}}
	\caption{The MSE of the proposed likelihood function estimate vs. data block index $n$ when adopting the ML detection method in $2\times 8$ and $4\times 8$ MIMO systems with one-bit ADCs.} \vspace{-3mm}
	\label{fig:LF_error}
\end{figure}

Fig.~\ref{fig:LF_error} plots the average MSE of the proposed likelihood function estimate, computed as
\begin{align}
\frac{1}{2N_{\rm rx}K}\sum_{i=1}^{2N_{\rm rx}}\sum_{k=1}^{K} |p_{i,k}-\hat{p}_{i,k}^{\sf pro}({\bf S})|^2,
\end{align}
versus the index of the data block when applying the ML detection. As a performance benchmark, the MSE in \eqref{eq:MSE_limit} derived for the ideal case of $\beta_k=1$ is also plotted. The parameters related to the transmission frame are set to be $N_{\rm p}=8N_{\rm tx}$, $D=40$, and $N_{\rm d}=128$. As can be seen in Fig.~\ref{fig:LF_error}, the MSE of the proposed likelihood function estimate significantly decreases with the data block index. This result demonstrates that the mismatch in the likelihood function is effectively reduced by using the proposed method, while the amount of the reduction increases as the number of input-output samples increases. Furthermore, the mismatch reduction in the $2\times 8$ MIMO case is shown to be larger than that in the $4\times 8$ MIMO case, because the larger the number of possible symbol vectors, the smaller the number of input-output samples per each symbol vector. It is also shown that the difference between the simulated MSE and  the ideal MSE in \eqref{eq:MSE_limit} is smaller in the $2\times 8$ MIMO case than in the $4\times 8$ MIMO case. The underlying reason is that for the same per-bit SNR, the $2\times 8$ MIMO system is more reliable than the $4\times 8$ MIMO system; thereby, $\beta_k$ in the $2\times 8$ MIMO system is closer to one than that in the $4\times 8$ MIMO system. 
    
\begin{figure}
	\centering 
	{\epsfig{file=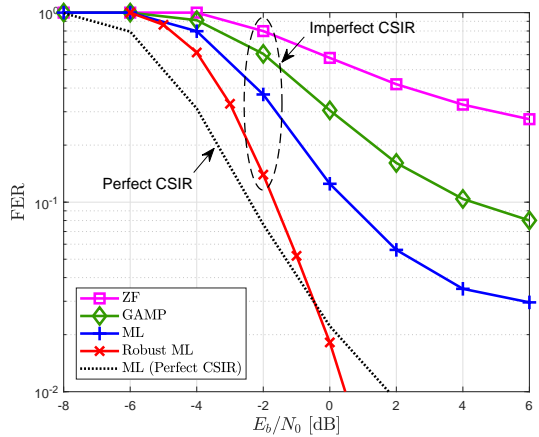, width=8cm}}
	\caption{The FER vs. $E_b/N_0$ of the ML, the robust ML, the GAMP-based, and the ZF detection methods for a time-varying frequency-flat channel in a $4\times 8$ MIMO system with one-bit ADCs.} \vspace{-3mm}
	\label{fig:FER_4x8_Time}
\end{figure}

Fig.~\ref{fig:FER_4x8_Time} compares the FERs of various detection methods for a time-varying frequency-flat channel in a $4\times 8$ MIMO system with one-bit ADCs. Particularly, we model the time-varying channel by adopting the first-order Gaussian-Markov process as done in \cite{Dong:04,Jeon:TCOM:16} which is a simple yet effective model to characterize the time-varying effect. Using this model, the channel matrix at time slot $n$ is obtained as 
\begin{align}
	{\bf H}^{(n)} = \sqrt{1-\epsilon^2}{\bf H}^{(n-1)} + \epsilon{\bm \Delta},
\end{align}
for $n\in\mathcal{N}_d$ and $d\in\{1,\ldots,D\}$, where ${\bf H}^{(0)}={\bf H}$, $\epsilon \in [0,1]$ is a temporal evolution coefficient, and each element of ${\bm \Delta} \in\mathbb{C}^{N_{\rm rx}\times N_{\rm tx}}$ is assumed to be independent and identically distributed as $\mathcal{CN}(0,1)$. We set $\epsilon=10^{-2}$ in the simulations. The parameters related to the transmission frame are set to be $N_{\rm p}=8N_{\rm tx}$, $D=10$, and $N_{\rm d}=128$ (i.e., 1024 coded bits). As can be seen in Fig.~\ref{fig:FER_4x8_Time}, when the channel varies over time, the ML detection with perfect CSIR suffers from the mismatch in the likelihood functions due to the channel variations. Furthermore, the performance loss of the ML detection with imperfect CSIR is even more severe in time-varying channels, as this method suffers from both the channel variations and the channel estimation error. Whereas, the ML detection using the proposed method is robust to both effects, because in the proposed method, any change in the likelihood function can be tracked by exploiting the input-output samples that \textit{empirically} provide the information of such change. Other suboptimal methods (GAMP-based detection and ZF detection) are still inferior to both the robust and the conventional ML detection methods in terms of the FER performance, as similar to Fig.~\ref{fig:FER_4x8_Flat}.

\subsection{Frequency-Selective Channels}
We also present simulation results for time-invariant frequency-selective channels. In this simulation, we consider the following data detection methods:
\begin{itemize}
	\item {\em Q-BCJR \cite{Jeon:arXiv:18}:} The optimal MAP detection method for wideband (frequency selective) MIMO systems with few-bit ADCs, which performs the BCJR algorithm based on the likelihood functions;
	\item {\em Q-BP \cite{Jeon:arXiv:18}:} A near-optimal low-complexity MAP detection method for wideband MIMO systems with few-bit ADCs, which performs the belief propagation (BP) algorithm based on the likelihood functions;
	\item {\em OFDM-Convex \cite{Studer:16}:} A joint-subcarrier data equalization method for MIMO-OFDM systems with few-bit ADCs, which solves a convex optimization problem using the FASTA algorithm;
	\item {\em OFDM-Bussgang:} A per-subcarrier data equalization method for MIMO-OFDM systems with few-bit ADCs, which linearizes the quantized received signal based on Bussgang's theorem \cite{Bussgang} under the assumption of the Gaussian signaling; and
	\item {\em OFDM-MMSE:} A per-subcarrier data equalization method for conventional MIMO-OFDM systems, which ignores the quantization effect at the ADCs (i.e., by assuming ${\bf y}[n]={\bf r}[n]$).   
\end{itemize}
We refer to the Q-BCJR method and the Q-BP method operating with the proposed likelihood learning method as \textit{robust Q-BCJR} and \textit{robust Q-BP}, respectively. For the channel estimation, we adopt a time-domain LMMSE method with $N_{\rm p}$ pilot signals which ignores the quantization effect at the ADCs. 

\begin{figure}
	\centering 
	{\epsfig{file=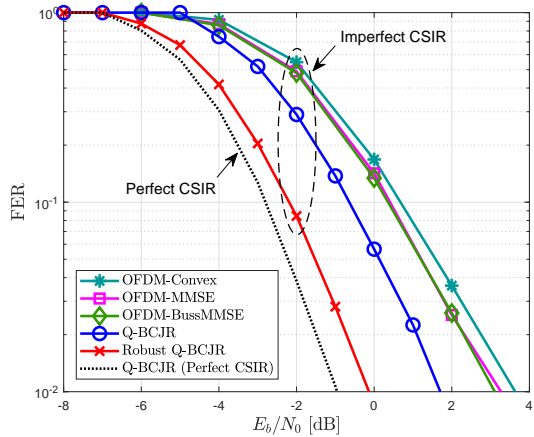, width=8cm}}
	\caption{The FER vs. $E_b/N_0$ of the Q-BCJR, the robust Q-BCJR, and the existing OFDM-based methods for a frequency selective channel in an $1\times 4$ SIMO system with one-bit ADCs.} \vspace{-3mm}
	\label{fig:FER_1x4_QBCJR}
\end{figure}

Fig.~\ref{fig:FER_1x4_QBCJR} compares the FERs of various detection methods for a frequency selective channel in an $1\times 4$ single-input multiple-output (SIMO) system with one-bit ADCs. In this simulation, the channel is modeled by independent Rayleigh fading CIR taps that follow an exponentially-decaying power-delay profile with an exponent 0.5 and $L=3$. The parameters related to the transmission frame are set to be $N_{\rm p}=8L$, $D=20$, and $N_{\rm d}=512$ (i.e., 1024 coded bits). Fig.~\ref{fig:FER_1x4_QBCJR} shows that the Q-BCJR with perfect CSIR achieves the lowest FER which is the optimal performance in the considered system. When this method is employed under imperfect CSIR, however, a significant performance loss is observed due to the mismatch in the model-based likelihood function. Whereas, in the robust Q-BCJR, the use of the proposed learning method effectively reduces this loss by correcting the likelihood function mismatch from the learning. As a result, the robust Q-BCJR is superior to all the other detection methods under imperfect CSIR. The FERs of OFDM-based methods are severely degraded not only by the channel estimation error, but also by the use of the OFDM signaling, as reported in \cite{Jeon:arXiv:18}.

\begin{figure}
	\centering 
	{\epsfig{file=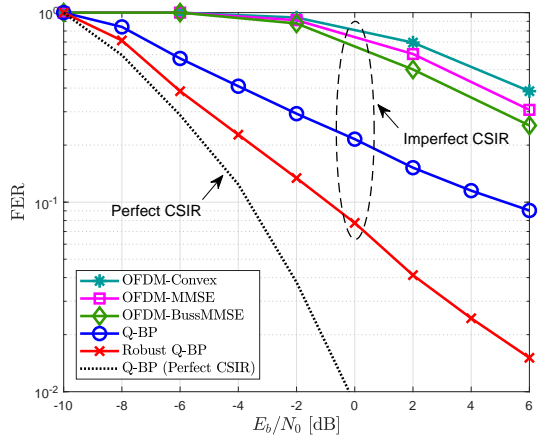, width=8cm}}
	\caption{The FER vs. $E_b/N_0$ of the Q-BP, the robust Q-BP, and the existing OFDM-based methods for a frequency-selective mmWave channel in an $1\times 8$ SIMO system with one-bit ADCs.} \vspace{-3mm}
	\label{fig:FER_1x8_QBP}
\end{figure}

Fig.~\ref{fig:FER_1x8_QBP} compares the FERs of various detection methods for a frequency-selective mmWave channel in an $1\times 8$ SIMO system with one-bit ADCs. In this simulation, the mmWave channel is implemented\footnote{In this implementation, the system bandwidth is set to be 1 GHz, the transmitter is assumed to use $4\times 4$ uniform-planar-array (UPA) with $N_{\rm tx}=1$ RF chain, and the receiver is assumed to use $8\times 8$ UPA with $N_{\rm rx}=8$ RF chains. The antenna-element spacing in both the horizontal and the vertical domains of the UPA is set to be $0.5\lambda$. The transmit and receive analog BFs are designed based on Algorithm~1 in \cite{Mo:BF:17}.} according to the 28-GHz non-line-of-sight model in \cite{Samimi:16}. Particularly, only the channels with a less than 4 dominant (more than $1\%$ of total power) CIR taps are used for simulations, in order to maintain an affordable level of the computational complexity when applying the Q-BP detection method\footnote{The Q-BP is applied with the dominant-tap-selection algorithm in \cite{Jeon:arXiv:18} with $D_{\rm max}=4$ and $\epsilon_{\rm th}=0.1$.}. The parameters related to the transmission frame are set to be $N_{\rm p}=3L$, $D=20$, and $N_{\rm d}=512$ (i.e., 1024 coded bits). For the channel estimation with $E_b/N_0\geq 0$ dB, we use the estimated channel obtained at $E_b/N_0\!=\!0$ dB, to prevent from a performance degradation caused when applying the LMMSE method in a high-SNR regime. As can be seen in Fig.~\ref{fig:FER_1x8_QBP}, the Q-BP with perfect CSIR achieves the lowest FER, as it is a near-optimal detection method for a mmWave MIMO system with one-bit ADCs \cite{Jeon:arXiv:18}. In the imperfect CSIR case, the robust Q-BP shows a substantial FER gain over the conventional Q-BP, which is attained by  using the proposed learning method. Particularly, this performance gain is shown to be larger in mmWave channels than in other wireless channels, due to a high channel estimation error in the mmWave channels. 

\section{Conclusion}
In this paper, we have presented a likelihood function learning method which is universally applicable to data detection methods that utilize the likelihood functions as the sufficient statistics in MIMO systems with one-bit ADCs. The key idea of the presented method is to exploit input-output samples obtained from the data detection, to improve the accuracy of likelihood function estimates. Inspired by the resemblance between the presented method and reinforcement learning, we have optimized this method by solving a reinforcement learning problem. One prominent feature is that the mismatch in the likelihood function decreases with the number of the input-output samples exploited in the presented method; this feature has been demonstrated by both the analysis and the numerical results. Using simulations, we have also shown that the use of the presented method makes the existing data detection methods robust not only to the channel estimation error but also to the effect of the channel variations.

A simple yet powerful extension of this work is to apply our approach for a communication system with time-varying channels. In this extension, our approach can be optimized to correct a mismatch effect caused by the channel variations. Another important direction for future research is to extend our approach for a communication system with hardware impairments beyond one-bit ADCs, in which the proposed approach can be used to correct a modeling error caused by hardware imperfections or imperfect knowledge of the system model at the receiver. When the knowledge of the system model is completely absent at the receiver, it would also be possible to develop a model-free communication framework by combining our approach with a supervised learning approach developed in our previous work \cite{Jeon:TVT:18}.

\appendices
\section{Proof of Lemma~1}\label{Apdx:Lem1}
Suppose that ${\bf S}=({\bf U},{\bf W},n)$ with $n\in \mathcal{N}_d$. From the definitions of the reward function ${\sf R}({\bf S},{\bf S}^\prime)$ and the optimal state ${\sf S}_{j,d}^\star({\bf S})$, ${\sf V}^\star\big({\sf S}_{j}^\prime({\bf S})\big)$ in \eqref{eq:Q_value} is expressed as 
\begin{align}\label{eq:Apdx:V_star}
	{\sf V}^\star\big({\sf S}_{j}^\prime({\bf S})\big) = {\sf R}\big({\sf S}_{j}^\prime({\bf S}),{\sf S}_{j,d}^\star({\bf S})\big).
\end{align}
Applying \eqref{eq:T_hat} and \eqref{eq:Apdx:V_star} into \eqref{eq:Q_value} yields 
\begin{align}\label{eq:Apdx:Q_value}
	{\sf Q}({\bf S},a)
	&= \sum_{j=0}^K {\sf T}_{a}\big({\sf S}_{j}^\prime({\bf S}),{\bf S}\big)
	\left\{ {\sf R}\big({\bf S},{\sf S}_{j}^\prime({\bf S})\big) + {\sf R}\big({\sf S}_{j}^\prime({\bf S}),{\sf S}_{j,d}^\star({\bf S})\big)\right\} \nonumber \\
	&\overset{(a)}{=} \sum_{j=0}^K {\sf T}_{a}\big({\sf S}_{j}^\prime({\bf S}),{\bf S}\big) \sum_{i=1}^{2N_{\rm rx}} \sum_{k\in\mathcal{K}} 
	\left\{{\sf MSE}_{i,k}^{\star}\big({\bf S}\big) 
	-  {\sf MSE}_{i,k}^{\star}\big({\sf S}_{j,d}^\star({\bf S})\big) \right\},
\end{align}
where (a) is obtained from \eqref{eq:Reward}. By removing terms in \eqref{eq:Apdx:Q_value} that are irrelevant to an action $a$, the optimal policy in \eqref{eq:Policy_opt0} is expressed as
\begin{align}\label{eq:Apdx:Policy_opt}
	\pi ({\bf S}) 
	&= \argmin_{{\bf a} \in [0,1] }~ \sum_{i=1}^{2N_{\rm rx}} \sum_{k\in\mathcal{K}} \sum_{j=0}^K {\sf T}_{a}\big({\sf S}_{j}^\prime({\bf S}),{\bf S}\big)
		{\sf MSE}_{i,k}^{\star}\big({\sf S}_{j,d}^\star({\bf S})\big) \nonumber \\
	&\overset{(b)}{=}  \argmin_{{\bf a} \in [0,1] }~ \sum_{i=1}^{2N_{\rm rx}} \sum_{j=0}^K {\sf T}_{a}\big({\sf S}_{j}^\prime({\bf S}),{\bf S}\big)
	 	{\sf MSE}_{i,\hat{k}[n]}^{\star}\big({\sf S}_{j,d}^\star({\bf S})\big), 
\end{align}
where (b) holds because  ${\sf MSE}_{i,k}^{\star}\big({\sf S}_{j,d}^\star({\bf S})\big)$ does not depend on an index $j$ for $k\neq \hat{k}[n]$ and $i\in\{1,\ldots,2N_{\rm rx}\}$.

\section{Proof of Theorem~1}\label{Apdx:Thm1}
By applying the approximate transition function in \eqref{eq:T_hat} into \eqref{eq:Policy_pro}, the optimal policy ${\pi} ({\bf S})$ is expressed as
\begin{align}\label{eq:Apdx:Policy_pro}
	{\pi} ({\bf S}) &=\mathbb{I}\left[\sum_{i=1}^{2N_{\rm rx}} \sum_{j=1}^{K} 
		\theta_{n,j} {\sf MSE}_{i,\hat{k}[n]}^{\star}\big({\sf S}_{j,d}^\star({\bf S})\big) - {\sf MSE}_{i,\hat{k}[n]}^{\star}\big({\sf S}_{0,d}^\star({\bf S})\big)
	\right].
\end{align}
To further characterize ${\pi} ({\bf S})$, we find a closed-form expression for the minimum MSE of the $(i,k)$-th likelihood function for given ${\bf w}_k$, denoted by ${\sf MSE}_{i,k}^{\star}({\bf S})=\argmin_{\alpha \in [0,1]}{\sf MSE}_{i,k}({\bf S};\alpha)$. For this, we rewrite the MSE for given $\alpha$ in \eqref{eq:MSE_def}  as
\begin{align}\label{eq:Apdx:MSE}
	&{\sf MSE}_{i,k}({\bf S};\alpha) 
	= \alpha^2 A_{i,k} + 2\alpha(1-\alpha) B_{i,k} + (1-\alpha)^2 C_{i,k} \nonumber \\ 
	&= \left(A_{i,k}+C_{i,k} -2B_{i,k} \right) 
	\!\left( \alpha -\frac{C_{i,k} -B_{i,k}}  {A_{i,k}+C_{i,k} -2B_{i,k}} \right)^{\!2} 
	\!+ \frac{A_{i,k}C_{i,k} -B_{i,k}^2}{A_{i,k}+C_{i,k} -2B_{i,k}},
\end{align}
where
\begin{align}
	A_{i,k} &= \mathbb{E}_{\{p_{i,k}\}_k,\hat{p}_{i,k}^{\sf emp}}\big[ (p_{i,k}-\hat{p}_{i,k}^{\sf mod})^2 ~\big|~{\bf w}_k \big],   \\
	B_{i,k} &= \mathbb{E}_{\{p_{i,k}\}_k,\hat{p}_{i,k}^{\sf emp}}\big[ (p_{i,k}-\hat{p}_{i,k}^{\sf mod}) (p_{i,k}- \hat{p}_{i,k}^{\sf emp}) ~\big|~{\bf w}_k \big],   \\
	C_{i,k} &= \mathbb{E}_{\{p_{i,k}\}_k,\hat{p}_{i,k}^{\sf emp}}\big[ (p_{i,k}- \hat{p}_{i,k}^{\sf emp} )^2 ~\big|~{\bf w}_k \big].	 
\end{align}
To compute three arguments $A_{i,k}$, $B_{i,k}$, and $C_{i,k}$, we specify the statistical characteristic of the empirical likelihood function $\hat{p}_{i,k}^{\sf emp}$ for given ${\bf w}_k$. Since 
\begin{align}
	\mathbb{P}\big[\tilde{y}_i[n]=+1\big] =\mathbb{P}\big[{y}_i^{\sf Re}[n]=+1\big]= p_{i,k[n]},
\end{align}
$\tilde{y}_i[n]$ is a Binomial random variable with mean $p_{i,k[n]}$ and one trial. Therefore, $u_{i,k}$ is a Poisson Binomial random variable whose mean and variance are given by
\begin{align*}
	\mathbb{E} [u_{i,k}] 
	&= \sum_{m\leq n}p_{i,k[m]}\mathbb{I}\{\hat{k}[m]=k\},  \\
	{\sf Var}(u_{i,k}) &= \sum_{m\leq n}p_{i,k[m]}(1\!-\!p_{i,k[m]})\mathbb{I}\{\hat{k}[m]=k\},
\end{align*}
respectively. The definition of $w_{j,k}$ in \eqref{eq:w_def} allows us to express the mean and the variance of the $(i,k)$-th empirical likelihood function as
\begin{align}
	\mathbb{E} \big[\hat{p}_{i,k}^{\sf emp}\big] 
	&= \frac{\sum_{j} w_{j,k}p_{i,j}}{\sum_j w_{j,k}} = \mu_{i,k}^{\sf emp}, \label{eq:Apdx:mu_emp} \\
	{\sf Var}\big(\hat{p}_{i,k}^{\sf emp}\big) 
	&= \frac{\sum_{j} w_{j,k}p_{i,j}(1\!-\!p_{i,j})}{(\sum_j w_{j,k})^2}  = \mathcal{E}_{i,k}^{\sf emp},  \label{eq:Apdx:var_emp}
\end{align}
respectively. Under the assumptions of $\mathbb{E}[p_{i,k}]=\hat{p}_{i,k}^{\sf mod}$ and ${\sf Var}({p}_{i,k})=\hat{\mathcal{E}}_{i,k}^{\sf mod}$ along with \eqref{eq:Apdx:mu_emp} and \eqref{eq:Apdx:var_emp}, three arguments $A_{i,k}$, $B_{i,k}$, and $C_{i,k}$ in \eqref{eq:Apdx:MSE} are computed as follows:
\begin{align}
	A_{i,k} &= \hat{\mathcal{E}}_{i,k}^{\sf mod},  \label{eq:Apdx:A_compute} \\
	B_{i,k} &= (1-\beta_k)\hat{\mathcal{E}}_{i,k}^{\sf mod},  \label{eq:Apdx:B_compute} \\
	C_{i,k} &= (1-2\beta_k) \hat{\mathcal{E}}_{i,k}^{\sf mod} + \frac{1}{\bar{u}_{k}^2}\Omega_{i,k}^{({\bf w}_k)},  \label{eq:Apdx:C_compute}
\end{align}
where $\beta_k=\frac{w_{k,k}}{\bar{u}_k}$, $\bar{u}_k=\sum_j w_{j,k}$,
\begin{align*}	
	\Omega_{i,k}^{({\bf w}_k)} &= \bigg(\sum_{j\neq k} w_{j,k} \big(\hat{p}_{i,k}^{\sf mod} - \hat{p}_{i,j}^{\sf mod}\big)\bigg)^2 
		+ \sum_{j} w_{j,k} \hat{v}_{i,k} + \sum_{j}  w_{j,k}^2  \hat{\mathcal{E}}_{i,j}^{\sf mod},
\end{align*}
and $\hat{v}_{i,k} = \hat{p}_{i,j}^{\sf mod}(1\!-\!\hat{p}_{i,j}^{\sf mod}) - \hat{\mathcal{E}}_{i,k}^{\sf mod}$.
From \eqref{eq:Apdx:A_compute}--\eqref{eq:Apdx:C_compute}, it can be easily shown that
\begin{align}\label{eq:Apdx:MSE_positive}
	A_{i,k}+C_{i,k} -2B_{i,k} &= \frac{1}{\bar{u}_{k}^2}\Omega_{i,k}^{({\bf w}_k)}	\geq 0.
\end{align}
The results in \eqref{eq:Apdx:MSE} and \eqref{eq:Apdx:MSE_positive} imply that the minimum MSE for given ${\bf S}$ is obtained as
\begin{align}\label{eq:Apdx:MSE_compute}
	{\sf MSE}_{i,k}^{\star}({\bf S}) 
	&= \min_{\alpha \in [0,1]} {\sf MSE}_{i,k}({\bf S};\alpha)
	= \frac{A_{i,k}C_{i,k} -B_{i,k}^2}{A_{i,k}+C_{i,k} -2B_{i,k}} = \bigg(1 - \frac{  w_{k,k}^2 \hat{\mathcal{E}}_{i,k}^{\sf mod} }	{   \Omega_{i,k}^{({\bf w}_k)} } \bigg)\hat{\mathcal{E}}_{i,k}^{\sf mod},	
\end{align}
and the corresponding combining ratio is given by
\begin{align}\label{eq:Apdx:alpha_compute}
	\alpha_{i,k}^\star({\bf S}) 
	&= \argmin_{\alpha \in [0,1]} {\sf MSE}_{i,k}({\bf S};\alpha) = \frac{C_{i,k} -B_{i,k}}  {A_{i,k}+C_{i,k} -2B_{i,k}} = 1 - \frac{ w_{k,k}\bar{u}_k \hat{\mathcal{E}}_{i,k}^{\sf mod}  }     {   \Omega_{i,k}^{({\bf w}_k)}  }.	
\end{align}
Finally, by applying both \eqref{eq:Apdx:MSE_compute} and \eqref{eq:S_opt_hat} into \eqref{eq:Apdx:Policy_opt}, we obtain the result in \eqref{eq:Policy_pro}.

\end{document}